%% file: main.tex
\documentclass[twocolumn]{aastex63}
\usepackage{float}
\usepackage{graphicx}
\usepackage{amsmath}
\usepackage{natbib}
\usepackage{color}
\usepackage{verbatim}
\usepackage{mathtools}
\usepackage{outlines}
\usepackage{ulem}
\usepackage{totcount}
\usepackage{textcomp}
\usepackage{afterpage}
\usepackage{tikz}
\usepackage{booktabs}
\usepackage{needspace}

\newcolumntype{s}{!{\extracolsep{-7pt}}c!{\extracolsep{0pt}}}
\newcolumntype{t}{!{\extracolsep{-12pt}}c!{\extracolsep{0pt}}}
\newcolumntype{p}{!{\extracolsep{-3pt}\textit{}}c!{\extracolsep{0pt}}}
\newcolumntype{b}{!{\extracolsep{10pt}\textit{}}c!{\extracolsep{0pt}}}

\newcommand{\neII}{[Ne\,\textsc{ii}]}
\newcommand{\neIII}{[Ne\,\textsc{iii}]}
\newcommand{\cII}{[C\,\textsc{ii}]}
\newcommand{\TIR}{\ensuremath{\textrm{\scriptsize TIR}}}
\newcommand{\um}{\,\textmu m}

\shortauthors{Smercina \textit{et al.}}
\shorttitle{After The Fall}

\begin{document}

\title{After The Fall: Resolving the Molecular Gas in Post-Starburst Galaxies}

\author[0000-0003-2599-7524]{Adam Smercina}
\affiliation{Astronomy Department, University of Washington, Seattle, WA 98195, USA}
\author[0000-0003-1545-5078]{John-David T. Smith}
\affiliation{Ritter Astrophysical Research Center, University of Toledo, Toledo, OH 43606, USA}
\author[0000-0002-4235-7337]{K. Decker French}
\affiliation{Department of Astronomy, University of Illinois, 1002 W. Green St., Urbana, IL 61801, USA}
\affiliation{National Center for Supercomputing Applications, 1205 W. Clark St, Urbana, IL, 61801, USA}
\author[0000-0002-5564-9873]{Eric F. Bell}
\affiliation{Department of Astronomy, University of Michigan, Ann Arbor, MI 48109, USA}
\author[0000-0002-5782-9093]{Daniel A. Dale}
\affiliation{Department of Physics \& Astronomy, University of Wyoming, Laramie, WY 82071, USA}
\author[0000-0001-7421-2944]{Anne M. Medling}
\affiliation{Ritter Astrophysical Research Center, University of Toledo, Toledo, OH 43606, USA}
\affiliation{ARC Centre of Excellence for All Sky Astrophysics in 3 Dimensions (ASTRO 3D)}
\author[0000-0003-1991-370X]{Kristina Nyland}
\affiliation{U.S. Naval Research Laboratory, 4555 Overlook Ave. SW, Washington, DC 20375, USA}
\author[0000-0003-3474-1125]{George C. Privon}
\affiliation{National Radio Astronomy Observatory, 520 Edgemont Road, Charlottesville, VA 22903, USA}
\affiliation{Department of Astronomy, University of Florida, 211 Bryant Space Sciences Center, Gainesville, FL 32611, USA}
\author[0000-0001-7883-8434]{Kate Rowlands}
\affiliation{AURA for ESA, Space Telescope Science Institute, 3700 San Martin Drive, Baltimore, MD, USA}
\author[0000-0003-4793-7880]{Fabian Walter}
\affiliation{Max-Planck-Institut f\"{u}r Astronomie, K\"{o}nigstuhl 17, D-69117 Heidelberg, Germany}
\author[0000-0001-6047-8469]{Ann I. Zabludoff}
\affiliation{Department of Astronomy, University of Arizona, Steward Observatory, Tucson, AZ 85721, USA}

\correspondingauthor{Adam Smercina}
\email{asmerci@uw.edu}

\begin{abstract}
Post-starburst (PSB), or `E+A', galaxies represent a rapid transitional phase between major, gas-rich mergers and gas-poor, quiescent early-type galaxies. Surprisingly, many PSBs have been shown to host a significant interstellar medium (ISM), despite theoretical predictions that the majority of star-forming gas should be expelled in AGN- or starburst-driven outflows. To-date, the resolved properties of this surviving ISM have remained unknown. We present high resolution ALMA continuum and CO(2--1) observations in six gas- and dust-rich PSBs, revealing for the first time the spatial and kinematic structure of their ISM on sub-kpc scales. We find extremely compact molecular reservoirs, with dust and gas surface densities rivaling those found in (ultra-)luminous infrared galaxies. We observe spatial and kinematic disturbances in all sources, with some also displaying disk-like kinematics. Estimates of the internal turbulent pressure in the gas exceed those of normal star-forming disks by at least 2 orders of magnitude, and rival the turbulent gas found in local interacting galaxies, such as the Antennae. Though the source of this high turbulent pressure remains uncertain, we suggest that the high incidence of tidal disruption events (TDEs) in PSBs could play a role. The star formation in these PSBs' turbulent central molecular reservoirs is suppressed, forming stars only 10\% as efficiently as starburst galaxies with similar gas surface densities. ``The fall'' of star formation in these galaxies was not precipitated by complete gas expulsion or redistribution. Rather, this high-resolution view of PSBs' ISM indicates that star formation in their remaining compact gas reservoirs is suppressed by significant turbulent heating. \\
\end{abstract}

\section{Introduction}
\label{sec:intro}
Major mergers represent one of the most rapid evolutionary pathways available to galaxies. In the current merger paradigm, colliding gas is heated in shocks, which quickly dissipate and trigger both starburst events and significant gas inflow towards the central potential, forming compact molecular reservoirs \citep{barnes&hernquist1991,barnes&hernquist1996,renaud2015,sparre&springel2016}. Dense, rotating gas disks form rapidly from these nuclear reservoirs --- as found for example in nearby ultra-luminous infrared galaxies \citep[ULIRGs, e.g.,][]{sakamoto1999,scoville2017}. In contrast, the majority of isolated elliptical galaxies --- the predicted end-state of such major mergers --- lack the substantial cool gas and dust reservoirs found in star-forming galaxies (\citealt{fall1979,barnes1992,cappellari2013,young2014}; typical molecular gas--to--stellar mass fractions of 0.01--1\%). If starbursting systems are the progenitors of many such `red-and-dead' early-type galaxies, the rarity of observed `transitioning' galaxies suggests that their significant ISM transformation must have been rapid ($\sim$1\,Gyr; \citealt{barro2014,schawinski2014}). The details of this transition from molecular gas-rich to gas-poor remain unclear.

Post-starburst galaxies (PSBs) are a rare galaxy class existing between these two crucial evolutionary phases. Though cataloged using a number of different classification schema, they are generally defined as galaxies that have experienced a significant decline in star formation rate (SFR) from a previous peak. Classical PSBs --- `E(K)+As'  \citep{dressler1983,couch&sharples1987} --- are galaxies with little-to-no nebular line emission, but a uniquely dominant $\sim$several 100 Myr old stellar population --- only possible if star formation was shut off very rapidly. While some PSBs reside in galaxy clusters, most are found in the field, where they are often identifiable as major merger remnants in \textit{Hubble Space Telescope} (\textit{HST}) imaging \citep{zabludoff1996,yang2006,chandar2021}. Thus, these PSBs provide a unique window onto the rapid time-steps following a major merger. 

In the established picture from hydrodynamical simulations, galaxies' molecular fuel is expelled in the end stages of major mergers by powerful feedback from star formation and active galactic nuclei (AGN) prior to and during the PSB phase \citep{hopkins2006,hopkins2008,snyder2011}. In direct contrast to this prediction of bulk ISM expulsion, recent work has found that many PSBs host large reservoirs of molecular gas \citep{french2015,rowlands2015,alatalo2016b} and significant dust emission in the near- and far-infrared \citep[][]{alatalo2017,smercina2018,li2019}. These reservoirs are unlike `typical' galaxies' ISM, as they exhibit unusually bright pure rotational molecular hydrogen (H$_{2}$) emission \citep{smercina2018} --- indicative of a warm, turbulent molecular ISM. Though any interpretation has so far been limited by the relatively coarse physical resolutions of the underlying observations, star formation in these galaxies may be turbulently suppressed from forming stars efficiently \citep{smercina2018}. Supportive of their possibly inefficient star formation, these galaxies exhibit an observed deficit of dense (i.e. `pre-stellar') gas, from ALMA limits on their HCN and HCO$^{+}$\ emission \citep{french2018}.

\input{obstbl}

While these discoveries have significantly advanced our understanding of the intermediate stages in the merger-driven evolutionary pathway, given their local rarity and, thus, large typical distance of PSB samples ($>$100\,Mpc), it has not been possible to investigate the structure and kinematics of the ISM in gas-rich PSBs. Where does their remaining ISM reside and is it forming stars as efficiently\footnote{Where `efficiency' refers to the expected density of stellar formation for a given gas density (in units of yr$^{-1}$), rather than a quantity related to the galaxy's stellar mass.} as expected in more rapidly star-forming systems? This lack of detail contrasts significantly with galaxies at the beginning \citep[e.g.,][]{ueda2014,scoville2017} and end \citep[e.g.,][]{ledo2010} of the major merger sequence --- a gap that must be filled in order to test and refine our understanding of the physical processes driving this rapid evolutionary transition. 

In this paper, we present, for the first time, high-resolution ALMA observations of six PSB galaxies, in the CO(2--1) 230.538 GHz line and the adjacent 1.3\,mm continuum.

\section{Observations \& Reduction}
\label{sec:obs}
Our ALMA intermediate baseline observations were taken across Cycle 3 \&\ 4 --- for project \#'s 2015.1.00665.S and 2016.1.00980.S, respectively (see Table \ref{tab:obs}). The Cycle 3 observations were obtained throughout the period from December 2015 to July 2016, while the Cycle 4 observations were taken in July (one source) and October 2016. Observations were obtained in relatively compact configurations, with maximum baselines ranging from $\sim$300\,m--3\,km, for angular resolutions\footnote{Estimated at 220.19 GHz} of $\sim$1\farcs5 (Cycle 3) and $\sim$0\farcs2 (Cycle 4). Additionally, 4 of the 6 sources were supplemented with 7\,m compact array (ACA) observations. 

All observations were obtained in receiver Band 6, with 4 spectral windows; each spectral window was set to a full bandwidth of 1875 MHz. One of the four windows was centered on the redshifted $^{12}$CO(2--1) line (rest frequency 230.538 GHz) for each source, while the remaining three windows were dedicated to dust continuum emission. The line spectral windows were set to 3.90 MHz ($\sim$5.4 km\,s$^{-1}$) resolution. During both cycles the three continuum-dedicated windows for each source were set to fixed 31.25 MHz ($\sim$40 km\,s$^{-1}$) resolution.

The data were pipeline-calibrated and further reduced using the Common Astronomy Software Applications package (CASA; \citealt{casa}) version 5.1.0. Where available, the Cycle 3 12\,m, Cycle 4 12\,m, and ACA observations were combined in the \textit{UV}-plane in order to provide both high resolution and sensitivity to more diffuse/extended emission. Self-calibration was performed on one source (0480). Briggs weighting with Robust = 0.5 was adopted for all sources. We used the \texttt{tclean} task (based on the CLEAN algorithm; \citealt{hogbom1974}) to image the continuum, using the three dedicated continuum spectral windows. Further iterations of \texttt{tclean} were run on the continuum-subtracted data in the line-centered spectral windows, using hand-chosen apertures, until approximately uniform residual images remained.

We used tclean to image the continuum, using the three dedicated continuum spectral windows

\input{restbl}

\section{Results}
\label{sec:results}
Here we present the results of our multi-cycle ALMA campaign to resolve the ISM properties of six PSBs. We first provide estimates of the cold dust continuum emission at 1.3\,mm in \S\,\ref{subsec:cont} in each of the six galaxies, followed by the resolved properties of their CO(2--1) emission (\S\,\ref{subsec:co-emission}), including their spatial (\S\,\ref{subsubsec:compact}) and kinematic (\S\,\ref{subsubsec:kin}) structure. Derived properties such as 1.3\,mm flux densities, 2-D elliptical Gaussian fits to the CO emission, and integrated CO line luminosities are given in Table \ref{tab:res}. Also given in Table \ref{tab:res} are ratios of the CO(2--1) integrated luminosity measured in this work to IRAM single-dish measurements. We discuss these ALMA/single-dish comparisons at the end of \S\,\ref{subsubsec:kin}. 

The six PSBs in this sample were originally identified in SDSS, but have since been assigned unique identifiers through follow-up study. Throughout this paper we refer to the PSBs by their SDSS plate numbers --- e.g., 0379, 0480, 0570, 0637, 2360, and 2777. The full plate-fiber-mjd identifiers published in \cite{smercina2018} can be found in Table \ref{tab:obs}, and their corresponding numerical designations adopted by \cite{french2015} are listed in Table \ref{tab:res}.

\begin{figure*}
\leavevmode
\centering
\includegraphics[width=0.85\linewidth]{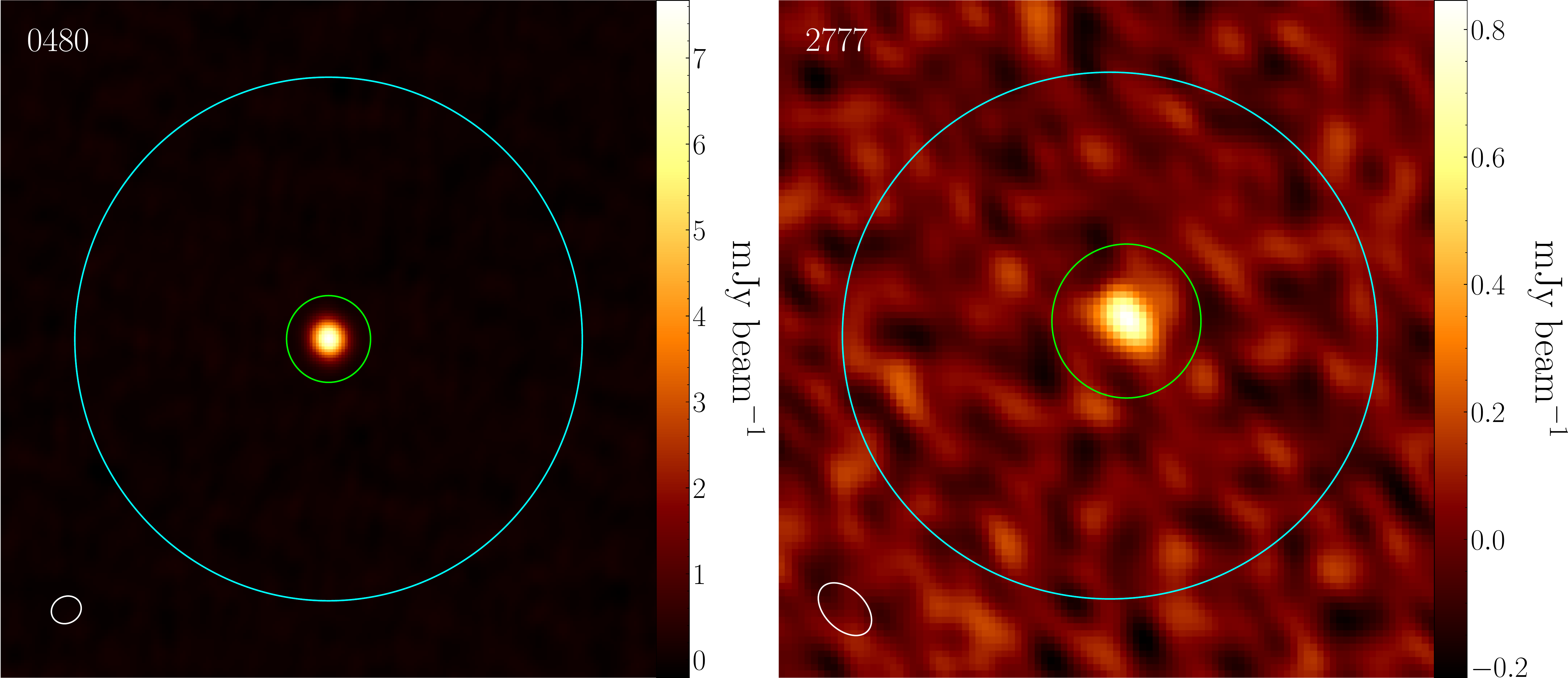}
\caption{Images of the 1.3\,mm continuum emission in our two galaxies with continuum detections, 0480 and 2777. The green circle corresponds to the 3$\sigma$\ apertures used to estimate the continuum flux for each galaxy. The cyan circle corresponds to the stellar half-light radius for each galaxy, derived from the Petrosian magnitude; 1\farcs66 and 3\farcs08, respectively. The white ellipse represents the reconstructed resolving beam for each observation; the continuum emission is unresolved in both cases. Images are shown at a linear stretch, with min/max scaling.}
\label{fig:continuum}
\end{figure*}

\subsection{Continuum Emission}
\label{subsec:cont}
\begin{figure*}[!ht]
\centering
\leavevmode
\includegraphics[width={0.95\linewidth}]{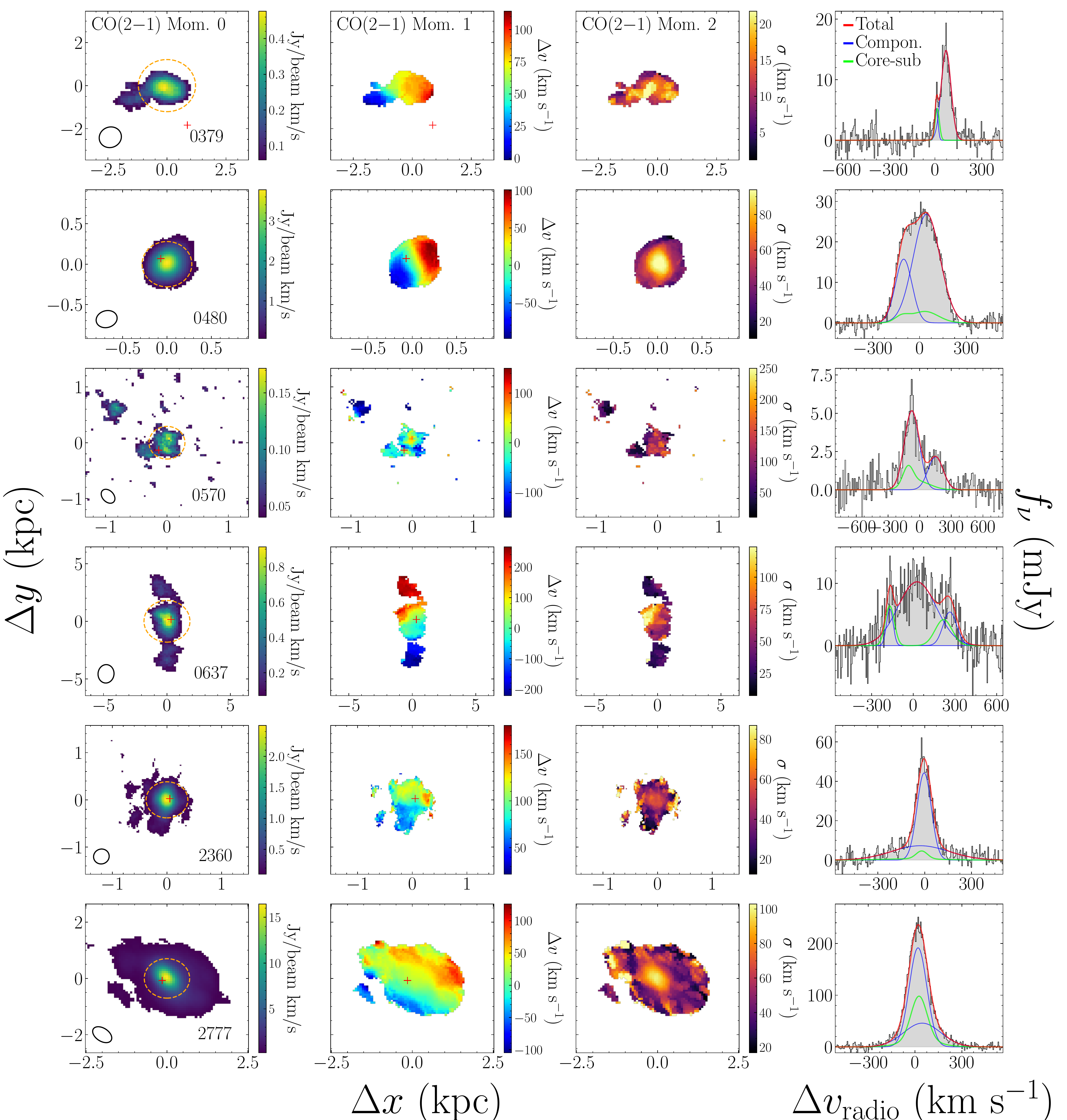}
\caption{Morphology and kinematics of the compact, turbulent CO(2--1) emission in our six PSBs observed with ALMA. \uline{Column 1}: Moment 0 images of CO(2--1) emission for each source. Images are centered on the CO cores and are oriented at PA = 0. The optical centers (from SDSS) are shown as red crosses. For all except 0379, offsets from the optical center are insignificant relative to the resolution of SDSS. The primary beam for each source is shown in the bottom left corner and the source ID in the bottom right. The `core' of each source is denoted by a dashed orange circle with radius equal to 3$\times$\ the average width (i.e. 3$\sigma$) of the 2D elliptical Gaussian fit (see Table \ref{tab:res}). \uline{Column 2}: Moment 1 intensity-weighted velocity maps. Velocity scale is centered on the systemic velocity of the host galaxy, measured from the SDSS spectrum. \uline{Column 3}: Moment 2 intensity-weighted velocity dispersion maps. \uline{Column 4}: Full line velocity profiles, extracted from the velocity fields shown in Column 2. The multiple Gaussian component fits to the profile are shown in blue and the total fit in red. A total fit to the velocity profile after subtracting the central core region is also shown, in green (see \S\,\ref{subsubsec:kin} for details).}
\label{fig:maps}
\end{figure*}

Measurements of the cold dust continuum at 1.3\,mm were performed on the continuum images described in \S\,\ref{sec:obs}. For sources with 1.3\,mm detections, apertures were chosen by-eye. The 1.3\,mm apertures coincide with the optical center of the galaxy in all detected cases (see Figure \ref{fig:maps}). For non-detections, continuum apertures were chosen based on detected CO emission sizes (see \S\,\ref{subsubsec:compact}), and limits were evaluated within these apertures at the 3$\sigma$\ level (where $\sigma$\ is the image RMS). Flux densities at 1.3\,mm, and corresponding uncertainties, are given in Table\,\ref{tab:res}.

We compare the 1.3\,mm continuum to model fits to the infrared (IR) spectral energy distributions (SEDs; 3--500\um), presented in \cite{smercina2018}. These fits use the method of \cite{dalehelou2002}, applied to the dust models of \cite{draineli2007}. The four non-detections, as well as 2777, are in good agreement with the best-fit IR SED models. 0480, however, displays an excess of emission at 1.3\,mm --- $\sim$3$\times$\ greater than expected from the best-fit SED. Far-infrared and sub-mm excess emission has been observed in nearby galaxies, such as in low-metallicity dwarf irregulars \citep[e.g.,][]{dale2012} and in unique galactic environments heavily populated by polarized, rotating dust grains \citep[e.g.,][]{murphy2010}. However, 0480 is unlikely to be low-metallicity, and 1.3\,mm is significantly higher energy than the rotating dust grain emission. This sub-mm excess also does not follow the expectation for free--free emission from obscured star formation, as 0480's radio emission (3 GHz from the VLASS survey \citealt{lacy2020}; 1.4 GHz from the FIRST survey \citealt{becker1995}) lies $\sim$2.5$\times$\ \textit{below} the radio--infrared correlation --- a statistically significant deficit, given the claimed scatter in the radio-IR correlation of 0.26\,dex, or a factor of $\sim$1.8 \citep[e.g.,][]{yun2001,bell2003}. We show the full infrared--radio SED in \textsc{Appendix} \ref{sec:radio-ir}. Whatever the source of this `anomalous' emission, it is not likely to be due to star formation, and does not significantly affect the estimate of 0480's TIR luminosity.

We show the continuum images for 0480 and 2777 in Figure \ref{fig:continuum}. Both galaxies were observed with sub-kpc physical resolution ($\sim$220\,pc and $\sim$550\,pc, respectively). In both cases 1.3\,mm continuum emission emanates from a point source, indicating that the dust reservoirs are highly compact. To estimate their dust column density, we assume that the 1.3\,mm emission is associated with their considerable global dust masses (\citealt{smercina2018}; likely, given the consistency with their global infrared SEDs). As both 0480's and 2777's continuum emission is unresolved, we divide their dust masses \citep[calculated in][]{smercina2018} by the size of the resovling beam (see Table \ref{tab:res}), yielding dust mass surface densities, $\Sigma_{\rm d}$, of 3200\,$M_{\odot}\,{\rm pc^{-2}}$\ and of 527\,$M_{\odot}\,{\rm pc^{-2}}$, respectively --- comparable to (and even exceeding) the dust surface densities seen in nearby ULIRGs \citep[e.g., Arp 220;][]{scoville2017}. Following typical prescriptions \citep{predehl&schmitt1995,guver&ozel2009}, visual extinction can be estimated from the dust column density as,\vspace{-3pt}
\begin{equation}
    A_V \approx \frac{\Sigma_{\rm d}/\left<DGR\right>}{2\times10^{21}\ {\rm cm^{-2}}}.
\end{equation}
Adopting a Nearby Galaxy-like dust-to-gas ratio of 1\% \citep{sandstrom2013}, this gives $A_V$\ of ${\sim}2{\times}10^4$\ and $\sim$3300, respectively. Since any distributed dust emission would reduce the inferred column densities, we regard these as illustrative upper limits. Though illustrative, these estimates indicate that the dusty reservoirs in these PSBs are unusually compact.  

\subsection{CO(2--1) Emission}
\label{subsec:co-emission}

\begin{figure*}[t]
\centering
\leavevmode
\includegraphics[width={0.95\linewidth}]{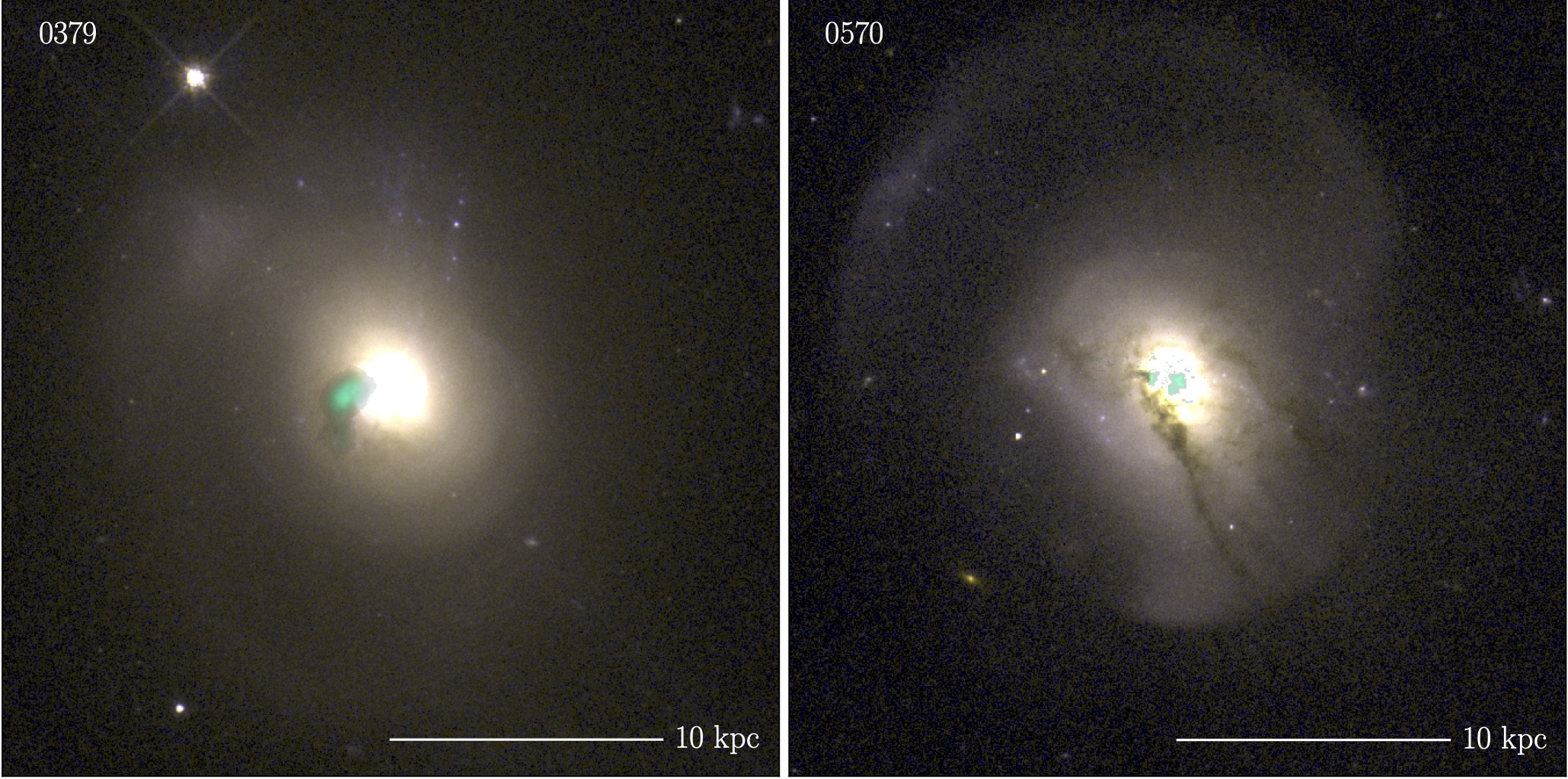}
\caption{Visual demonstration of the compactness of the PSBs' CO emission. CO emission in 0379 (left) and 0570 (right), is overlaid in green on archival \textit{HST} WFC3 F438W/F625W images. Images are shown in the \textit{HST} reference frame, with roll angles of 64\textdegree\ and 27\textdegree\ (counterclockwise), respectively. Images were scaled using a square root stretch and are displayed at equivalent physical scale. In both cases, the CO emission is highly compact relative to the optical extent. 0379's is off-set from the optical center, coincident with a faint dust lane. Both galaxies' gas exhibits a disturbed morphology, but with bright cores hosting the majority of the emission.}
\label{fig:hst}
\end{figure*}

\subsubsection{Morphology}
\label{subsubsec:compact}

In Figure \ref{fig:maps} we show the CO(2--1) Moment 0 integrated intensity, Moment 1 intensity-weighted velocity, and Moment 2 intensity-weight velocity dispersion maps for the six PSBs. Pixels were clipped at 3.5$\times$\ the channel RMS following computation of the moments of the spectral cube. The CO emission in all six sources is highly compact. Sizes were determined by fitting 2-dimensional elliptical Gaussian functions to each Moment 0 image. All six sources exhibit unresolved `cores' --- the FWHMs recovered from the elliptical fits are equivalent to the beam dimensions. The fraction of emission in the unresolved core varies for each source, ranging from 46\% in 0570 --- in which the emission arises from three individual compact clumps --- to $\gtrsim$95\% in 0480.

Figure~\ref{fig:hst} shows the CO emission of two PSBs (0379 \& 0570) on the same spatial scale as \textit{HST} observations of their starlight. The CO overlaid on the WFC3 F438W/F625W images (GO 11643; PI Zabludoff) highlights the compactness of the molecular gas relative to the galaxies' optical extents, as well as its disturbed morphology. In one of these sources, 0379, the CO emission is offset from the optical center, and is instead coincident with a faint dust lane. 

\begin{figure*}[t]
\centering
\leavevmode
\includegraphics[width={0.75\linewidth}]{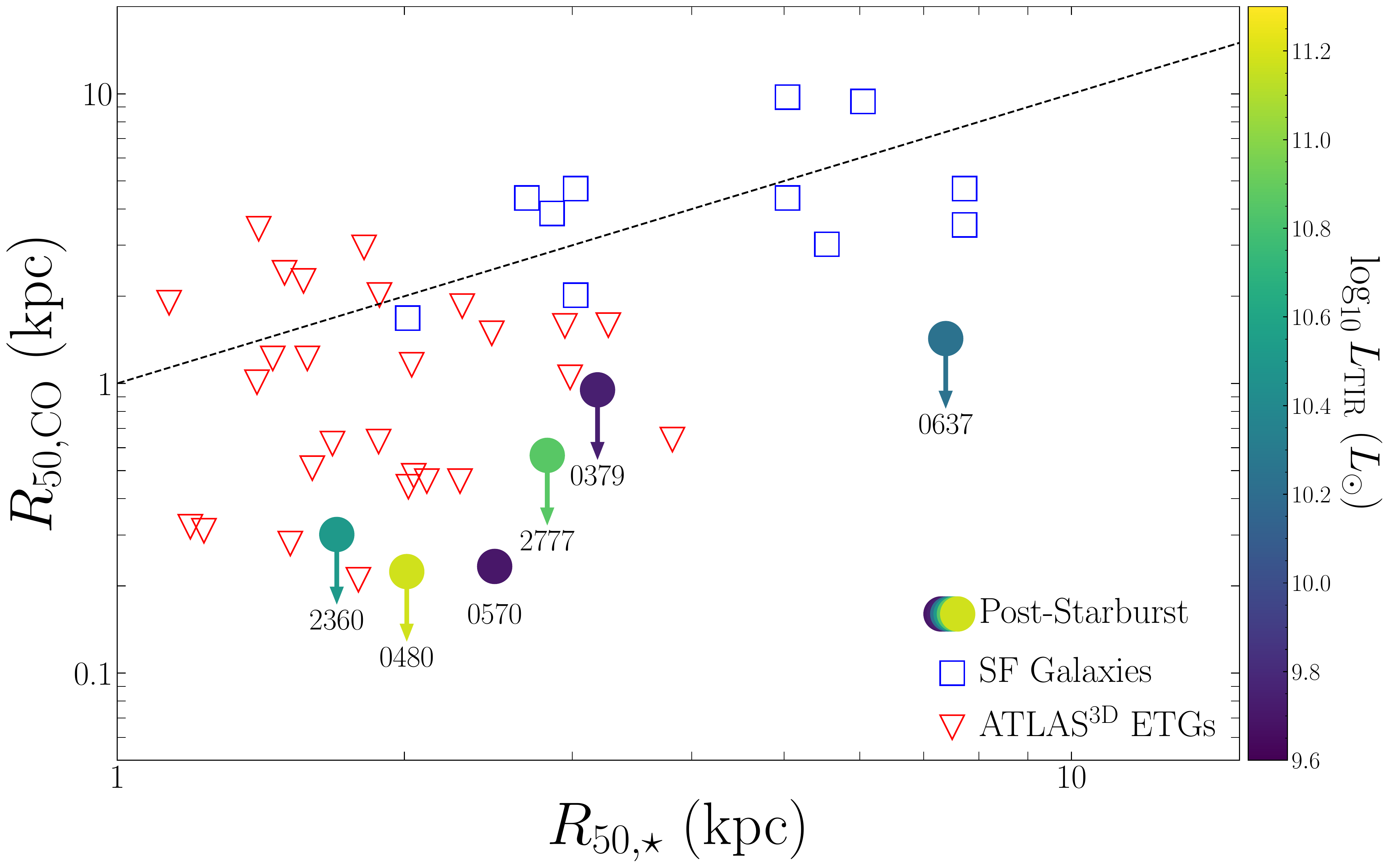}
\caption{Comparison of the stellar and CO sizes of the PSBs and other galaxy samples. Stellar half-light radius \citep[from SDSS imaging;][]{french2015,smercina2018} is plotted against CO half-light radius for the six PSBs (filled circles; this work), color-coded by their total infrared (TIR) luminosity. The PSBs are compared against a sample of nearby star-forming galaxies \citep{regan2006}, and early-type galaxies from the ATLAS$^{\rm 3D}$\ survey \citep{davis2013}. The CO half-light radii for ATLAS$^{\rm 3D}$\ galaxies are all upper limits (these sizes reflect the `maximum extent' of the CO emission). The dashed line shows a one-to-one relation, which the star-forming comparison sample approximately follow. $R_{\rm 50,CO}$\ for the PSBs is formally an upper limit (marked by downward arrows) for five of the six, as their core fractions ($f_{\rm core}$) are $>$50\% and our ability to probe the 50\% radius is limited by resolution --- i.e. the observed 50\% radius is governed by the beam profile, rather the physical emissive region. On average the CO in our PSB sample is ${>}5{\times}$\ more compact than the normal galaxy sample, despite stellar half-light radii similar to the normal galaxy sample. Roughly half of the ATLAS$^{\rm 3D}$\ early-type galaxies exhibit upper limits on their CO sizes comparable in compactness to upper limits on the PSBs.} 
\label{fig:sizes}
\end{figure*}

Though resolution and global morphology are not uniform across the sample, most of the CO emission in the sample is compact and found on sub-kpc scales. In Figure \ref{fig:sizes}, we compare the approximate CO half-light radii of these six PSBs to that of their stellar emission, and contrast this measure of gas-to-stellar compactness with a sample of nearby star-forming galaxies \citep[from][]{regan2006} and early-type galaxies (ETGs) from the ATLAS$^{\rm 3D}$\ survey \citep{davis2013}. The molecular gas in our PSB sample is, on average, $>$5$\times$\ more compact than that of a comparable sample of `normal' star-forming galaxies, both in absolute scale and relative to their stellar emission. The compactness of the PSBs' gas is more comparable to the most compact ETGs, though with $\sim$10$\times$\ the gas masses.

\subsubsection{Kinematics{\label{subsubsec:kin}}}
In Figure \ref{fig:maps} (right column), we show CO(2--1) velocity profiles for the six galaxies. Profiles were extracted from apertures drawn to match the shapes of the emission in the clipped Mom.\ 0 maps shown in Figure \ref{fig:maps}. Aperture selection and spectral extraction was performed using CASA. Profiles were fit with up to three Gaussian components. Also shown are profiles with the central `core' subtracted --- i.e. the `core' region is defined using a circular radius equal to 3$\times$\ the average width of the 2-D elliptical Gaussian fits (shown as orange ellipses in the left column of Figure \ref{fig:maps}). As the bulk of the gas in five of the six galaxies is spatially unresolved (\S\,\ref{subsubsec:compact}), we do not perform any sophisticated disk modeling; our kinematic analysis is limited here to identification of multiple components. The results for each of the six galaxies are summarized below. 

\uline{0379}: Best-fit by a two-component model, with the smaller component centered at the galaxy's systemic velocity. The larger component corresponds entirely with the CO `core', and is both spatially and kinematically off-center with respect to the galaxy's optical center by $\sim$2\,kpc and $\sim$71\,km\,s$^{-1}$.

\uline{0480}: A distinctive `double-peak' profile, typically suggestive of rotation. The CO in this system is almost entirely in the `core' --- it is unresolved at the achieved $\sim$200\,pc physical resolution. The unresolved nature of the emission precludes us from classifying it as a `rotating disk', though the gas certainly possesses high angular momentum content. 

\uline{0570}: Best-fit with two components, roughly symmetric around the systemic velocity. It is unlikely that this indicates rotation, however, as 0570's CO field is highly asymmetric, with three spatially distinct `clumps'. The core, and brightest, of these clumps is co-spatial with the optical center. The second brightest of the clumps appears to overlap with a dust lane.

\uline{0637}: Best-fit with three components --- one central, broad component, and two symmetric `wings' red- and blue-shifted by $\sim$200\,km\,s$^{-1}$. These symmetric wings correspond directly to the two spatially symmetric patches of emission to the North and South of the central core (see Figure \ref{fig:maps}), each extending several kpc along the galaxy's minor axis. These emission wings are oriented perpendicular to 0637's visible major axis in SDSS. We tentatively characterize these features as consistent with outflowing gas. The gas velocity is quite slow ($\sim$200\,km\,s$^{-1}$) compared to the fast AGN- and star formation-driven outflows seen at high-redshift \citep[e.g.,][]{sell2014}, and may be more comparable to local `slow-boil' winds, such as in the nearby galaxy M82 \citep[e.g.,][]{leroy2015}. This could also be a relic of past nuclear feedback. 

\uline{2360}: A strong central peak is recovered, along with a faint, broader component. The majority of the emission is located within the central core, with some faint extended `patchy' emission at larger radii. Both the velocity peak and broad underlying component appear to originate in the central core, while the kinematics of the extended emission mirror the central peak velocity profile. There does not appear to be strong rotation. 

\uline{2777}: Similar kinematics to 2360, with a central peak and broader underlying component. As with 2360, the majority of the emission is within the central core, with slightly more extended emission at larger radii. Both the velocity peak and broad underlying component again appear to originate in the central core, while the kinematics of the extended emission mirror the central peak velocity profile. There is no indication of strong rotation. 

All six galaxies, with the exception of the completely unresolved 0480, exhibit `patchy' structure in their velocity dispersion maps. The average velocity dispersions are substantially higher at these physical scales than in `typical' nearby galaxies \citep[e.g.,][on comparable physical scales]{sun2018}, with most of the gas in all six galaxies exhibiting $\sigma_{\rm disp}\,{\gg}$\,10\,km\,s$^{-1}$\ and up to 100\,km\,s$^{-1}$\ in some cases. 0637 is the only galaxy with distinct emission peaks perpendicular to the galaxy that we identify as a potential outflow. Both 2360 and 2777 exhibit broad velocity components associated with more spatially extended emission, which could be suggestive of outflowing gas \citep[e.g.,][]{}. Though without higher-resolution observations this remains highly speculative.

\subsection{Molecular Gas Mass and Surface Density}
\label{subsec:mmol}

We calculate integrated CO(2--1) luminosities, following \cite{solomon1997}, as
\begin{small}
\begin{equation}
    L_{\rm CO}^{\prime} = 3.25{\times}10^7\,(1{+}z)^{-3}\,\nu_{obs}^{-2}\,\,S_{\rm CO}\,D_{\rm L}^2, 
\end{equation}
\end{small}
where $z$\ is the redshift, $\nu_{obs}$\ is the observed frequency of the line in GHz, $D_{\rm L}$\ is the luminosity distance in Mpc \citep[from][assuming the \citealt{planck-cosmo} cosmology]{smercina2018}, and $S_{\rm CO}$\ is the integrated line flux in Jy\,km\,$s^{-1}$, calculated by integrating the multi-component velocity profile fits (red curves in Figure \ref{fig:maps}). 

We compare our integrated CO(2--1) line luminosities to the estimates in \cite{french2015}, which presented IRAM single-dish observations of both CO(1--0) and CO(2--1). The ratio of the IRAM measurement to our corresponding ALMA measurement is given in Table \ref{tab:res}. Our inclusion of 7-m ACA observations was intended to recover any emission on large scales captured by the single-dish observations that may be resolved out by our high-resolution 12-m observations. For three of the six galaxies the integrated CO(2--1) fluxes from ALMA and IRAM are relatively consistent --- they formally agree within 2$\sigma$. 0637 is slightly discrepant at 2.5$\sigma$, however we find that its IRAM spectrum is particularly noisy and is statistically consistent with the best fit velocity profile to the ALMA data. The remaining two galaxies, 0379 and 0480, are discrepant at $>$3.5$\sigma$, with significantly more flux estimated for 0379 with IRAM and significantly less flux for 0480 (0480 is formally a non-detection with IRAM). We rule out the possibility that this discrepancy in 0379 is due to resolving out extended emission in the ALMA observations, as the IRAM beam is 11\arcsec\ and the largest angular scale (LAS) of our ALMA observations is 19\farcs4 due to the inclusion of the 7-m ACA. This galaxy is one of the most marginal detections of the \cite{french2015} sample, and the disagreement may be due to uncertainties in the flux calibration between the samples, or poor continuum subtraction. A flux calibration or continuum subtraction issue seems the only likely explanation for the `missing' flux in the IRAM observations for 0480. In summary, these ALMA observations presented here appear to be broadly consistent with the existing single-dish observations, but their are some subtleties associated with comparing the two samples due to the order-of-magnitude higher signal-to-noise of these observations. We note that if the flux for any of these sources were indeed higher (to match the IRAM measurement), this would only increase the high molecular gas densities we observe for these objects, and would not affect our qualitative conclusions.

We convert the calculated CO(2--1) luminosities, $L_{\rm CO(2{-}1)}^{\prime}$, to total mass of molecular gas, $M_{\rm Mol}$, assuming a Galactic CO-to-H$_2$\ conversion factor\footnote{Including a 1.36 factor for Helium \citep[e.g.,][]{sandstrom2013}.} of $\alpha_{\rm CO}$\ = 4.35 M$_{\odot}$\,(K km s$^{-1}$\ pc$^{2}$)$^{-1}$\ and a $^{12}$CO(2--1)/CO(1--0) line ratio, $R_{21} = 0.59$\ in K\,km\,s$^{-1}$\ units \citep[most current value for nearby galaxies;][]{denbrok2021}. These new masses are given in Table \ref{tab:res}. $R_{21}$\ is relatively well constrained, with uncertainties on average at the $<$30\% level (though it likely varies from galaxy-to-galaxy). There is also uncertainty in the assumed $\alpha_{\rm CO}$\ conversion factor. However, \cite{smercina2018} found good general agreement between (1) molecular gas masses derived from H$_2$\ pure rotational emission and CO(1--0) using a Galactic CO-to-H$_2$\ conversion, as well as (2) the estimated dust-to-molecular gas ratios of the PSB sample and dust-to-gas ratios of nearby galaxies --- both suggesting that the adopted $\alpha_{\rm CO}$\ is an appropriate conversion factor for the PSB sample. 

Using these newly-calculated masses, and our estimates of the spatial scale of the CO emission, we can estimate the PSBs' molecular gas surface densities. We report the molecular gas surface densities, $\Sigma_{\rm Mol}$, within these central cores --- noting that in the majority of cases this captures the majority of the emission ($>$63\% in all but one galaxy; see Table \ref{tab:res}).

All six sources possess extremely high central molecular column densities --- $\Sigma_{\rm Mol}\,{=}\,344{-}6.5{\times}10^{4}\,M_{\odot}\,{\rm pc^{-2}}$\ --- the highest of which are consistent with the high inferred dust column densities measured in 0480 and 2777 (\S\,\ref{subsec:cont}). Unlike nearby galaxies, the entire gas reservoirs of these PSBs' exist on $<$\,1\,kpc scales, with surface densities that are therefore orders-of-magnitude higher than their nearby galaxy counterparts ($\sim$300--8$\times$10$^4$\ vs.\ $\sim$3--50\,$M_{\odot}\,{\rm pc^{-2}}$; e.g., \citealt{bigiel2008}) --- instead rivaling the gas found in the densest compact starbursts \citep[e.g.,][]{kennicutt&delosreyes2021}.

\section{Suppression of Star Formation in a Turbulent ISM}
\label{sec:ks}
Previous works have compared PSBs' global SFRs to their global CO emission \citep{french2015,smercina2018,french2018b}, making assumptions about the physical scales of each. Without proper measurements of these physical scales, it has not been possible to reliably estimate quantitative properties of their molecular gas reservoirs, such as the internal turbulent pressure, or how efficiently these reservoirs are forming stars.

In this section we first estimate the internal turbulent pressure for a subset of our sample and discuss comparisons to nearby galaxies (\S\,\ref{subsec:turb}), followed by an analysis of the suppression of the PSBs' star formation, relative to the star formation law (\S\,\ref{subsec:sf-law}). 

\subsection{Turbulent Pressure}
\label{subsec:turb}
Of the six PSBs studied in this work, three were observed at particularly high physical resolution. The beam sizes for 0480, 0570, and 2360 are only a few hundred parsecs --- comparable to the `driving scale' for interstellar turbulence \citep{brunt2003}, and approaching the scale of individual GMCs. The internal turbulent pressure for an individual cloud is proportional to its surface density, $\Sigma$, and line-of-sight velocity dispersion, $\sigma_v$, as $P_{\rm turb} \sim \Sigma\sigma_v^2$. As the molecular gas is concentrated on scales comparable to the resolving beam, we can estimate $P_{\rm turb}$\ for the `cores' of these three galaxies, with the approximation derived by \cite{sun2018}:
\begin{small}
\begin{equation}
    P_{\rm turb}/k_B \approx 61.3\ {{\rm K\,cm^{-3}}} 
    {\left( \frac{\Sigma_{\rm Mol}}{M_{\odot}\,{\rm pc^{-2}}} \right)}
    {\left( \frac{\sigma_v}{{\rm km\,s^{-1}}} \right)^2}
    {\left( \frac{R_{\rm beam}}{40\ {\rm pc}} \right)^{-1}},
\label{eq:3}
\end{equation}
\end{small}
\hspace{-0.45\parindent} where $\Sigma_{\rm Mol}$\ is the surface density of the molecular core described in \S\,\ref{subsubsec:compact}, $\sigma_v$\ is the average velocity dispersion measured in the core, and $R_{\rm beam}$\ is the radius of the beam (averaged over both dimensions) given in Table \ref{tab:res}. For the cores of 0480, 0570, and 2360, we estimate $\log_{10}P_{\rm turb} [{\rm K\,cm^{-3}}] =$\ 9.8, 9.1, and 8.8, respectively. 

\begin{figure}[t]
\centering
\leavevmode
\includegraphics[width=0.9\linewidth]{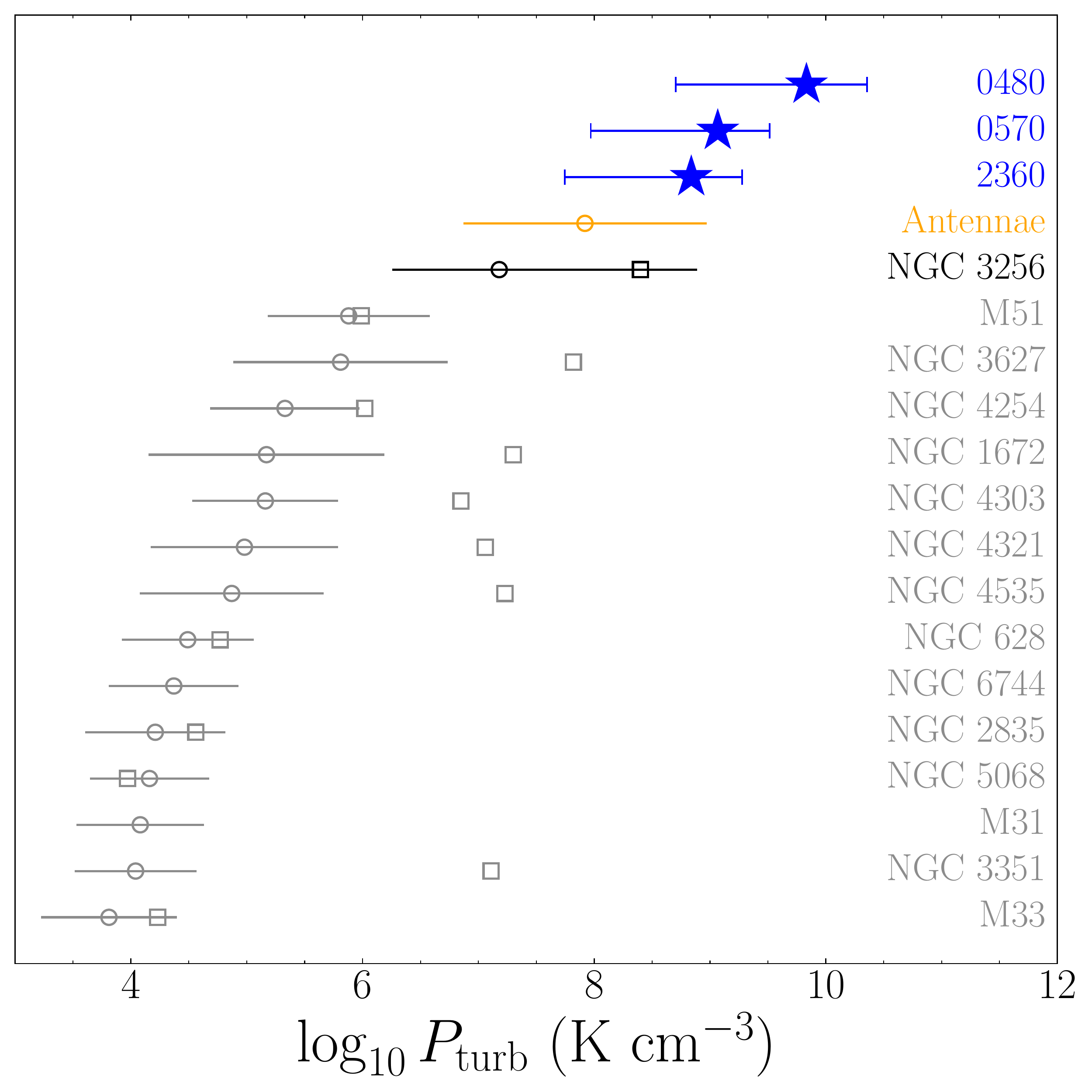}
\caption{Comparing the high turbulent pressures found in PSBs to known galaxies. Turbulent pressure ($P_{\rm turb}$) in the PSBs 0480, 0570, and 2360 (blue stars), compared against $P_{\rm turb}$\ for local galaxies from \cite{sun2018} and NGC 3256 \citep{brunetti2021}. Formal uncertainties on $P_{\rm turb}$\ for the PSBs are shown, incorporating the measured uncertainties on $\Sigma_{\rm Mol}$\ and $\sigma_v$. The median for the local galaxy disks are shown as gray circles, with the 16--84\% widths of the distribution of $P_{\rm turb}$\ across all resolved regions in each disk shown as gray lines, and the median for the nuclei (where available) as gray squares. NGC 3256 and the Antennae are shown separately, in black and orange, respectively. While too disturbed to be called a `disk', NGC 3256's `non-nuclear' average is shown as a black circle. Galaxies are ordered by increasing $P_{\rm turb}$\ along the y-axis.}
\label{fig:pturb}
\end{figure}

One limitation of our observations in drawing inferences about these high turbulent pressures is the possible smearing of high rotational velocities at scales below our physical resolution by the resolving beam. This `beam smearing' can add significant width to the measured velocity dispersion \citep[e.g., see][]{walter2022}. This a particular concern for sources with considerable rotational velocity signatures, such as 0480. To check the impact of beam smearing on our measured $\sigma_{v}$, we simulated rotating disks\footnote{Using the KinMS software package (Python version), written by Timothy Davis \citep{davis2013}}, comparable in size, shape, and velocity scale to 0480, with several combinations of different intrinsic gas velocity dispersions and rotation curves. We present the full analysis in \textsc{Appendix} \ref{sec:beamsmear}, but summarize the results here. In the case of a steeply-declining central component to the rotation curve, such as the presence of a central supermassive black hole, beam smearing could contribute substantially to the measured velocity dispersion at the scale of our observations. However, we find that beam smearing alone, with a much lower velocity dispersion, does not reproduce the observed Moment 2 image or velocity profile well. In the case of 0480 specifically, we find that a somewhat lower gas velocity dispersion of 50\,km\,s$^{-1}$\ (vs.\ 69\,km\,s$^{-1}$\ measured) and a 2$\times$10$^7\,M_{\odot}$\ central black hole reproduce the observed velocity profile fairly well (see Figure \ref{fig:maps}). This would imply an overestimation of $P_{\rm turb}$\ of 0.3\,dex, reducing it to 10$^{9.5}$\ --- a significant difference, but still much higher than the comparison sample in Figure \ref{fig:pturb}. We conclude that while observations do indicate much higher-than-average gas velocity dispersions, they are not high enough resolution to completely break this degeneracy between smeared rotation from a central point mass and high intrinsic velocity dispersion. We therefore introduce an additional factor of 10 lower uncertainty on $P_{\rm turb}$, to reflect the possibly significant contribution of beam smearing in our measured velocity dispersions. For context, this would place a lower limit on $\sigma_{v}$\ in 0480 of 40\,km\,s$^{-1}$, given its measured 69\,km\,s$^{-1}$. As shown in Figure \ref{fig:pturb}, even if $\sigma_v$\ was of order 20--30 km\,s$^{-1}$\ for PSBs, leading to a 10$\times$\ overestimate in turbulent pressure, their $P_{\rm turb}$\ values would still be higher than anything besides the Antennae (and higher still for 0480), due to their high surface densities. 

In Figure \ref{fig:pturb} we compare these estimates to the turbulent pressures measured by \cite{sun2018} in local galaxies, as well as the LIRG NGC 3256 \citep{brunetti2021}. The turbulent pressure in the three PSBs is 2--3 orders of magnitude higher than the median $P_{\rm turb}$\ measured for gas in the local star-forming disks, and is $>$10$\times$\ higher than found in local star-forming nuclei. The closest comparison in the local galaxy sample is the gas in the merging Antennae, which has a median $\log_{10}P_{\rm turb} = 7.9$, and the late-stage merger NGC 3256, which has a median $\log_{10}P_{\rm turb} = 7.2$\ in non-nuclear regions and $\log_{10}P_{\rm turb} = 8.3$\ in the nucleus. Even with possible contribution of beam smearing to their measured velocity dispersions, it is clear: these PSBs' molecular reservoirs are \textit{highly turbulent} relative to `typical' galaxies and rival (or even surpass) the turbulent pressures found in late-stage mergers such as the Antennae and NGC 3256. Higher-resolution observations with ALMA should refine these measurements even further, and allow more principled modeling of the rotation curves.

It is interesting to consider the physical interpretation of such high turbulent pressures in non-merging galaxies. The typical definition of turbulence is that it is part of a true energetic `cascade' down to small physical scales, where it must be accompanied by commensurate dissipation through mechanically heated emission channels. This seems to be at least partially supported by the observed bright pure rotational $H_2$\ emission, which is a hallmark of mechanically heated systems \citep[e.g.,][]{alatalo2014,appleton2017,smercina2018}. This scenario would likely favor a physical \textit{injection} of energy, such as local feedback from AGN activity, active on short enough timescales to counteract the efficient emissive dissipation. The modest outflow observed in the PSB 0637 could be a possible signature of such lower-level current or past AGN activity. However, if feedback alone was the source of turbulent energy injection we would expect to observe higher-energy cooling lines that are bright enough to dissipate the turbulent energy in the gas. An alternative mechanism is that the the apparent turbulent motions represent semi-coherent bulk flows of gas on scales smaller than the achieved physical resolution. This effect has been observed in Arp 220, and is thought to be driven primarily by merger-induced gravitational torques \citep{scoville2017}. The measured line-of-sight velocity dispersions for these three PSBs range from $\sim$47--112 km\,s$^{-1}$\ --- very comparable to the velocity dispersions measured in the two nuclei of Arp 220 \citep{scoville2017}.

It seems most likely that these two mechanisms work in concert. As discussed in \cite{scoville2017} in the nuclear disks of Arp 220, coherent bulk flows could result in an ISM that is more smooth than the `clumpy' ISM seen in normal galaxies. A largely smooth medium could explain the dearth of high density molecular tracers, like HCN and HCO$^+$\ \citep{french2018}, as it could be simultaneously consistent with the observed $H_2$\ rotational emission in PSBs and the lack of optical and infrared nebular cooling lines that would would typically drive rapid dissipation of turbulent energy and fragmentation of the gas. If we assume a very simple model where the molecular gas in the cores of 0480, 0570, and 2360 is spherically distributed, then we obtain average densities of $n$\,$\sim$\,10$^{3}{-}10^{4}$\ --- all below the effective critical density of HCO$^+$\ and well below that of HCN \citep{leroy2017,beslic2021}. If the ISM is as smoothly distributed as suggested by such a simple model, then the high surface densities could be entirely consistent with a relative lack of truly `dense gas'. It should be noted that dense gas tracers such as HCN have been observed in Arp 220's nuclei \citep{barcosmunoz2018}, despite a potentially `smooth' ISM, though this HCN emission is concentrated in outflows. We stress that this `toy model' view of a smooth ISM in PSBs is proposed as a \textit{possible} explanation only for their high gas surface densities, yet lack of truly dense gas. Upcoming studies with dense gas and CO excitation analysis will provide further in depth information on the state of their ISM \citep[e.g.,][]{french2022}.

\begin{figure*}[t]
\centering
\leavevmode
\includegraphics[width={0.9\linewidth}]{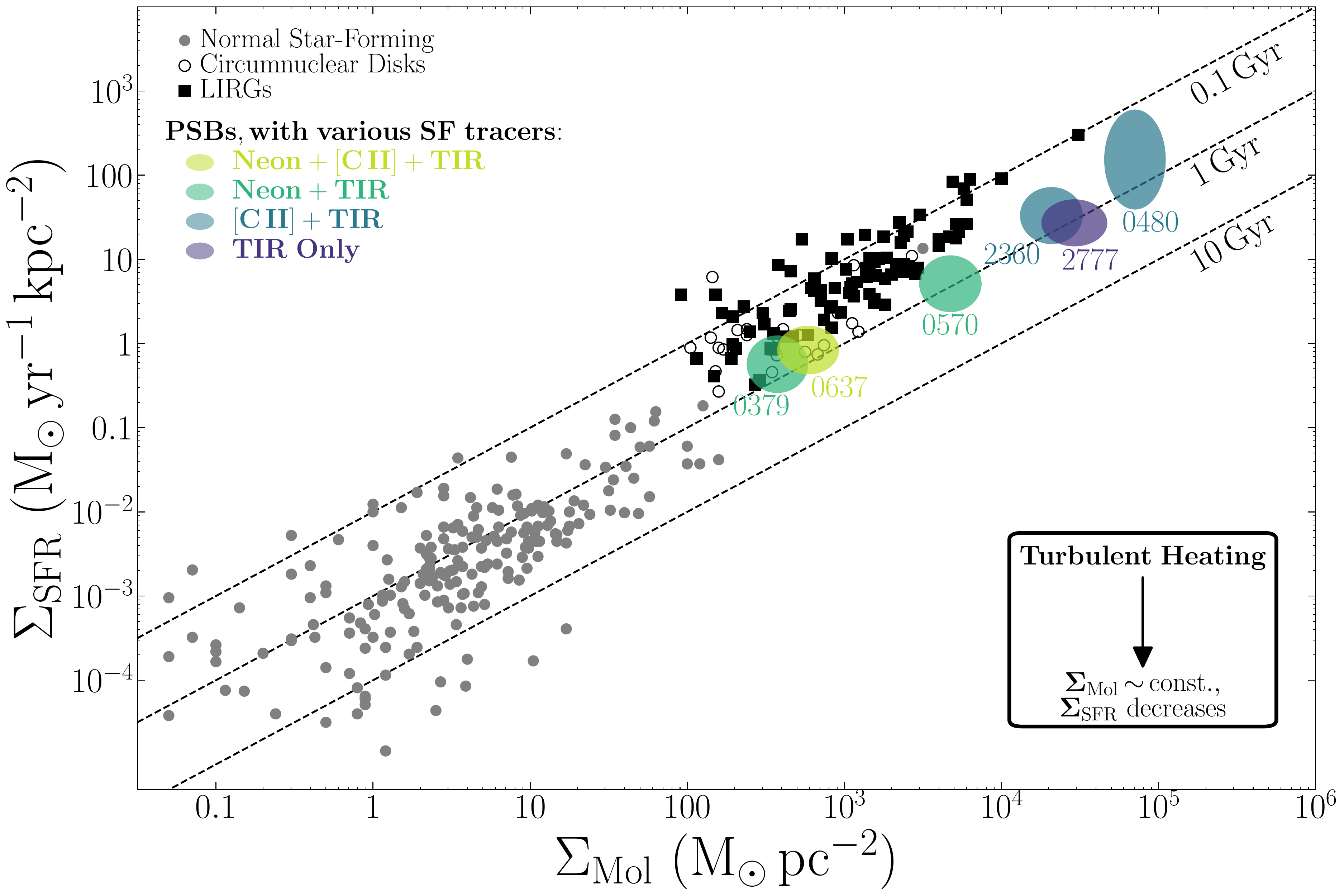}
\caption{Molecular gas and star formation of the PSB sample compared to the Kennicutt-Schmidt star formation relation. SFR surface density ($\Sigma_{\rm SFR}$) plotted as a function of molecular gas surface density ($\Sigma_{\rm Mol}$). The six PSBs are plotted as ellipses, color-mapped to four distinct combinations of the infrared SFR indicators used \citep{smercina2018}: Neon (\neII+\neIII), Neon + TIR, \cII + TIR, and TIR only --- corresponding to different combinations of existing observations from \textit{Spitzer} and \textit{Herschel}. The elliptical FWHMs and molecular gas masses listed in Table \ref{tab:res} are adopted and combined to calculate physical surface densities. The widths of each ellipse in $\Sigma_{\rm Mol}$\ and $\Sigma_{\rm SFR}$\ represent the uncertainties on each quantity (see \S\,\ref{subsec:sf-law}). Nearby galaxies are shown for comparison, drawn from \cite{delosreyes&kennicutt2019} and \cite{kennicutt&delosreyes2021}, divided into normal star-forming galaxies 
(gray filled circles), local circumnuclear disks (black open circles), and LIRGs (black filled squares). A uniform MW-like CO-to-$H_2$\ conversion factor ($\alpha_{\rm CO}$\ = 4.35) is assumed for all galaxies. Constant gas depletion timescales of 0.1, 1, and 10 Gyr are shown as dashed lines. With a factor of 10 improvement in spatial resolution over existing \textit{Spitzer} observations \citep{smercina2018}, the compactness of these molecular reservoirs results in substantial (5--10$\times$) offsets from the comparison sample. Though their molecular gas surface densities remain comparable to analogs of the PSBs own progenitor starbursts, the turbulent nature of these remaining reservoirs appears to have significantly reduced the efficiency of residual star formation.
}
\label{fig:ks}
\end{figure*}

Perhaps the best source of turbulent energy injection in these PSBs is some form of low-level low-duty cycle AGN feedback. Among the physical mechanisms for such low-level AGN feedback, an interesting point of consideration is the mounting evidence of a high incidince of tidal disruption events (TDEs) --- the disruption of a star by a central supermassive black hole and the corresponding energetic emission event --- in PSBs. TDEs occur $\sim$30--100$\times$\ more frequently in PSBs than in other galaxies \citep[e.g.,][]{french2016,stone&metzger2016,french2018c} which could be attributable to their high central stellar densities \citep[e.g.,][]{norton2001}. Compact central gas reservoirs, such as the high molecular gas column densities observed in the PSBs presented here, would only increase the gravitational drag on central stellar orbits, potentially supporting this picture.\footnote{Similarly, TDE rates are also expected to be high in starburst galaxies \citep{mattila2018,stone2018}, though they are much harder to observe due to obscuration by dust and contamination with AGN emission.} PSBs have no observed signatures of strong AGN feedback. However, high rates of TDEs could provide a form of consistent, low-level AGN feedback --- a possible avenue for generating their turbulent gas reservoirs --- while remaining consistent with the lack of AGN-tracing emission lines \citep[e.g.,][]{charalampopoulos2021,zabludoff2021}.

We consider here a simple model for TDE-generated feedback and compare it to the estimated turbulent energies observed. Assuming an average TDE rate typical in PSBs of $10^{-3}\,{\rm yr^{-1}}$\ \citep{french2016} and a typical integrated energy yield of $\sim$10$^{51}$\,erg for a UV/optical-detected TDE \citep{holoien2016}, TDEs could replenish the entire turbulent energy in the PSBs' compact molecular cores --- $\int{P_{\rm turb}}\,k_{\rm B}\ dV\,{\sim}\,10^{55}$\,erg --- in just 10\,Myr. This scenario incorporates two important assumptions. The first is that their ISM is mostly smooth (as discussed above). In a typical ISM, not ordered in smooth bulk flows, the turbulent dissipation timescale would be roughly the dynamical timescale at the scale of the gas --- $\sim$a few Myr for 0480, 0570, and 2360, assuming their measured velocity dispersions and CO size scales. However, even in a medium with typical nebular cooling lines, while it may take 10\,Myr for TDEs to replenish the PSBs' entire turbulent energy reservoirs, this TDE `duty cycle' could be short enough to counteract the turbulent dissipation of the gas. The second assumption is that the energy produced by the TDE can couple efficiently to the ISM. While this type of `feedback' has not yet been well-studied, recent models suggest that the radiative emission, which approximately follows a few$\times$10$^4$\,K blackbody, may be able to effectively couple to the cold surrounding medium, at least at distances very near the black hole \citep{bonnerot2021} --- similar to the large scales on which continuous black hole feedback is observed to operate. Though dedicated simulations are required to verify the plausibility of TDEs as a distinct source of black hole feedback, they will be an important mechanism to consider as the field attempts to understand the origin of the turbulent central reservoirs in these PSBs.

\subsection{The Star Formation Law}
\label{subsec:sf-law}

It has been shown previously that these PSBs have low global SFRs \citep{french2015,smercina2018,french2018,french2018b}, and we have shown here that their gas is highly compact and likely turbulent. This is important, as previous studies have found observational and theoretical evidence for the ability of turbulent energy injection to inhibit star formation \citep{alatalo2015,piotrowska2021}. How does the efficiency of the residual star formation in these compact, turbulent reservoirs compare to other known galaxies?

In Figure \ref{fig:ks}, we compare the surface density of molecular gas ($\Sigma_{\rm Mol}$) and star formation ($\Sigma_{\rm SFR}$) in our PSB sample to the Kennicutt--Schmidt star formation law taken from normal star-forming and compact infrared luminous galaxies \citep{kennicutt&delosreyes2021}. We use the quantities given in Table \ref{tab:res}, as well as SFR indicators based on the \neII\ \&\ \neIII\ mid-infrared fine structure lines, the \cII\ 158\,\um\ far-infrared cooling line, and the total 3--1100\,\um\ infrared (TIR) luminosity \citep{smercina2018}. Infrared SFR tracers are preferred in such dusty, compact systems, as they can penetrate all but the highest columns. Yet, they suffer from their own limitations. All galaxies have a reliable TIR from multi-band SED fitting, but the unique radiation fields in PSBs can cause the infrared emission to overestimate the current SFR \citep{hayward2014,smercina2018}. The \cII\ line, where detected, is similarly limited. \cite{smercina2018} found \neII+\neIII\ to be the most reliable indicator of the current SFR, though these lines are only available for the three galaxies observed with \textit{Spitzer} IRS.

All of the sources are unresolved and their SFR estimates are global. We therefore assume that star formation activity is spread over the same surface area as the 1.3\,mm continuum and CO emission. While star formation may be `clumpy' within the molecular gas region, we do not expect there to be substantial star formation outside of the region of molecular gas emission. We therefore calculate SFR surface density, $\Sigma_{\rm SFR}$, identically to $\Sigma_{\rm Mol}$: by using the size of the resolving beam, which corresponds to the unresolved core of each galaxy, and multiplying by the core fraction, $f_{\rm core}$.

There are numerous sources of uncertainty (largely systematic) in each of these metrics, and we conservatively estimate them from measurement uncertainty, uncertainty in the $\alpha_{\rm CO}$\ conversion factor, and uncertainty on the measured sizes of the molecular gas and star formation. We do not account for uncertainty in the different timescales that each indicator likely probes. Rather than fixed points with error bars, we chose to represent these uncertainties as error ellipses in Figure \ref{fig:ks}, with the width in each direction giving the likely spread in $\Sigma_{\rm Mol}$--$\Sigma_{\rm SFR}$\ parameter space. The width of the ellipse in $\Sigma_{\rm Mol}$\ is calculated by adding in quadrature the uncertainties on $L^{\prime}_{\rm CO(2{-}1)}$\ (see \S\,\ref{subsec:mmol} and Table \ref{tab:res}), a 50\% uncertainty on the CO size reflecting the possibility that the gas deviates from a Gaussian distribution as we have assumed, and a conservative 80\% uncertainty on the assumed $\alpha_{\rm CO}$\ ($\sim$0.8--7). The widths in $\Sigma_{\rm SFR}$\ include a 25\% uncertainty for a given SFR indicator \citep[following][]{smercina2018}, 50\% uncertainty on the star-forming size (following the CO), and spread from the average of the different indicators used. 

Even taking into account these uncertainties, Figure \ref{fig:ks} demonstrates the very significant suppression of star formation efficiencies of PSBs, relative to galaxies with similar gas densities. Their gas densities are 2--4 orders of magnitude higher than star-forming galaxies with similar global SFRs ($\sim$0.5--5 $M_{\odot}\,{\rm yr^{-1}}$), yet, despite possessing ULIRG-like surface densities, these reservoirs are forming stars only 10\% as efficiently, on average. Coupled with the high $P_{\rm turb}$\ found in \S\,\ref{subsec:turb}, we conclude that turbulent support in the ISM is a likely culprit for this star formation suppression. Turbulent suppression of star formation is  though it remains unclear over what timescales this turbulent suppression could persist or its driving mechanism. 

\section{Conclusions}
\label{sec:conclusions}
We have presented the results of high-resolution CO(2--1) ALMA observations of six post-starburst galaxies. For the first time, we have resolved the spatial distribution and kinematics of the ISM in this enigmatic class of galaxies. We find:
\begin{enumerate}
    \item Unresolved, and therefore highly compact, 1.3\,mm continuum emission in two galaxies (0480 and 2777). Inferred dust surface densities are extremely high --- $\Sigma_{\rm d}\,{=}$\,3200 and 527 $M_{\odot}\,{\rm pc^{-2}}$, respectively. These measured dust column densities correspond to visual extinctions of $A_V\,{\sim}\,2{\times}10^4$\ and $\sim$3300, respectively.
    \item Highly compact, largely centrally-concentrated molecular reservoirs. The majority of the emission in each source is contained within an unresolved core. The scale of the CO emission is much smaller than in typical star-forming galaxies, both in an absolute sense and relative to the extent of their stellar emission. We derive molecular gas surface densities ranging from 300 $M_{\odot}\,{\rm pc^{-2}}$\ to an incredible $7{\times}10^4\ M_{\odot}\,{\rm pc^{-2}}$\ --- comparable to those found in the most vigorously star-forming galaxies, such as ULIRGs.
    \item A large diversity in morphology and kinematic structure. Some galaxies' emission is highly patchy, with multiple distinct spatial components, while others' gas is entirely contained in the core. We see examples of clear rotation in one source and a possible moderate-velocity outflow in another.
    \item Large internal turbulent pressure ($P_{\rm turb}\,{>}\,10^8$\ K\,cm$^{-3}$) in the molecular gas of the three galaxies with the highest-resolution observations (0480, 0570, 2777). Though beam smearing of coherent motions may significantly contribute to the high measured velocity dispersions, the turbulent pressures in these PSBs are at least 2 orders-of-magnitude higher than found in normal star-forming galaxies, and likely higher even than in nearby mergers, such as the Antennae and NGC 3256. 
    \item We speculate that a smoothly-distributed ISM in these turbulent reservoirs may explain the high gas column densities, yet lack of observed dense-gas tracer emission. We further speculate that the high gas column densities could help to explain the recent finding of particularly high tidal disruption event (TDE) rates in PSBs, through increased orbital drag. We show that these high TDE rates could provide a plausible reservoir of constant energy injection, in the form of low-duty cycle AGN feedback, and could help to explain the high turbulent pressures in such a simple, smooth model ISM lacking typical AGN signatures.
    \item Star formation suppressed by a factor of $\sim$10$\times$\ relative to galaxies with comparable gas surface densities. While $\Sigma_{\rm Mol}$\ across the PSB sample is comparable to infrared luminous galaxies, $\Sigma_{\rm SFR}$\ is 10$\times$\ lower than in their starbursting counterparts. We assert that turbulent heating of these compact reservoirs directly results in this suppression of star formation efficiency. 
\end{enumerate}

Our results confirm that the ISM conditions and distribution in these PSBs are unlike normal galaxy populations. Understanding the mechanism driving the turbulent heating of their ISM, and whether this mechanism can maintain the low star-formation efficiency until the remaining gas is consumed, or rendered ineffective for star formation in a future evolutionary state, is of continued importance. The apparently paradoxical state of the gas and of star formation in these unique systems may provide a blueprint for understanding how star formation is quenched and regulated in galaxies `after the fall'. \\

We thank the anonymous referee for a careful and thoughtful review that improved this paper. A.S.\ was supported by NASA through grant \#GO-14610 from the Space Telescope Science Institute, which is operated by AURA, Inc., under NASA contract NAS 5-26555.  J.D.T.S. acknowledges visiting support from the Alexander von Humboldt Foundation and the Max Planck Institute f\"{u}r Astronomie. A.M.M acknowledges support from the National Science Foundation under Grant No. 2009416.

This paper makes use of the following ALMA data: ADS/JAO.ALMA\# 2015.1.00665.S and 2016.1.00980.S. ALMA is a partnership of ESO (representing its member states), NSF (USA), and NINS (Japan), together with NRC (Canada), NSC and ASIAA (Taiwan), and KASI (Republic of Korea), in cooperation with the Republic of Chile. The Joint ALMA Observatory is operated by ESO, AUI/NRAO, and NAOJ. We thank the ALMA support staff at NRAO Charlottesville --- particularly Sarah Wood and Anthony Remijan --- for invaluable help in the reduction and analysis of these observations. The National Radio Astronomy Observatory is a facility of the National Science Foundation operated under cooperative agreement by Associated Universities, Inc. Basic research in radio astronomy at the U.S. Naval Research Laboratory is supported by 6.1 Base Funding.

\facility{ALMA}
\software{\texttt{CASA} \citep{casa}, \texttt{Matplotlib} \citep{matplotlib}, \texttt{NumPy} \citep{numpy-guide,numpy}, \texttt{Astropy} \citep{astropy}, \texttt{SciPy} \citep{scipy}, \texttt{SAOImage DS9} \citep{ds9}, \texttt{KinMS} \citep{davis2013}}

\bibliographystyle{aasjournal}

\appendix

\vspace{-20pt}
\section{Infrared to Radio SED for 0480}
\label{sec:radio-ir}
\begin{figure}[!h]
    \centering
    \includegraphics[width=0.6\linewidth]{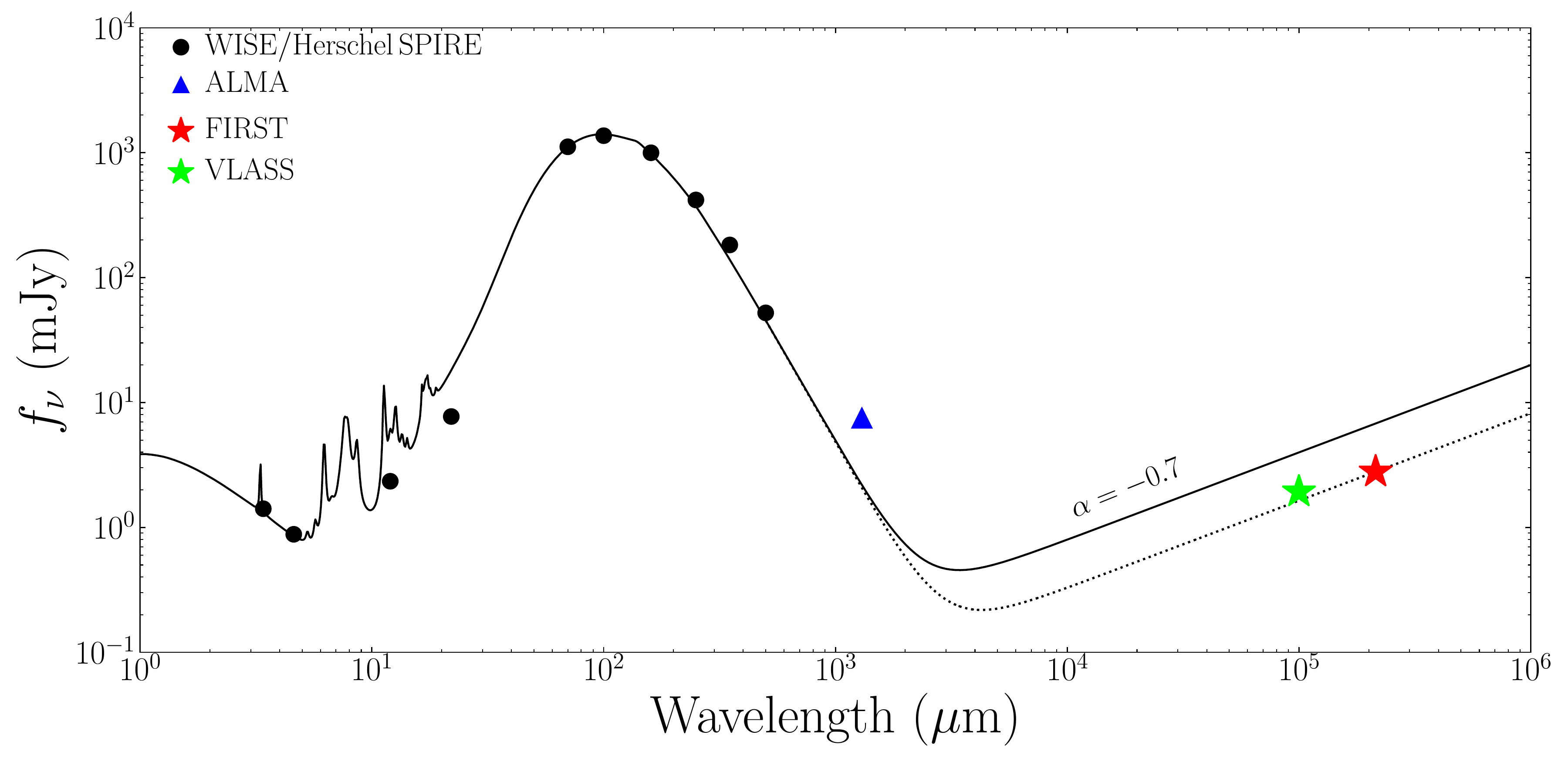}
    \caption{Infrared--radio SED for 0480, along with the best-fit infrared SED from dust models combined with the expected radio SED of slope $-$0.7 from the radio--infrared correlation (solid black line). An SED, with a $-$0.7 radio power-law 2.5$\times$\ below the radio--infrared expectation, is also shown as a best fit to the VLASS and FIRST data.}
    \label{fig:irradio-sed}
\end{figure}

Figure \ref{fig:irradio-sed} shows all existing photometry for 0480 from the infrared through the radio. \textit{WISE} and \textit{Herschel} SPIRE photometry is taken from \cite{smercina2018}, the ALMA 1.3\,mm flux is taken from this work (see \S\,\ref{subsec:cont}), and the radio comes from the VLASS (3\,GHz; \citealt{lacy2020}) and FIRST (1.4\,GHz; \citealt{becker1995}) surveys. We also show the best-fit model SED to the infrared photometry, following the method of \cite{dalehelou2002}, applied to the dust models of \cite{draineli2007} and presented in \cite{smercina2018}. The 1.4--3\,GHz radio SED is $\sim$2.5$\times$\ below the expectation from the radio--infrared correlation, given 0480's inferred TIR luminosity. The deviation of the ALMA measurement must then be due to an additional emission component, and not due to free-free emission from obscured star formation. Additional radio observations in the $\sim$33\,GHz range should provide clarity on the source of this anomalous emission.

\vspace{-20pt}
\section{Data for Figures 4, 5, and 6}
\label{sec:figure-data}

The derived quantities (sizes, velocity dispersions, densities, and turbulent pressures) presented in Figures \ref{fig:sizes}, \ref{fig:pturb}, and \ref{fig:ks} are given in Table \ref{tab:dens}.

\input{denstbl}

\clearpage
\section{The Effects of Beam Smearing on Measured Velocity Dispersions}
\label{sec:beamsmear}

Here we investigate the effect of beam smearing on measured velocity dispersions in the PSBs. We use 0480 as our standard for this analysis, as it is the only source that exhibits a clear rotational signature in its Moment 1 map and velocity profile. We use the KinMS Python package \citep{davis2013} to construct model gas disks similar to 0480, at the same resolution as the observations. Given that we observe rotation, we assume the disks are relatively highly inclined, with inclination $i$\,=\,65\textdegree.

\begin{figure}[!h]
    \centering
    \includegraphics[width=0.55\linewidth]{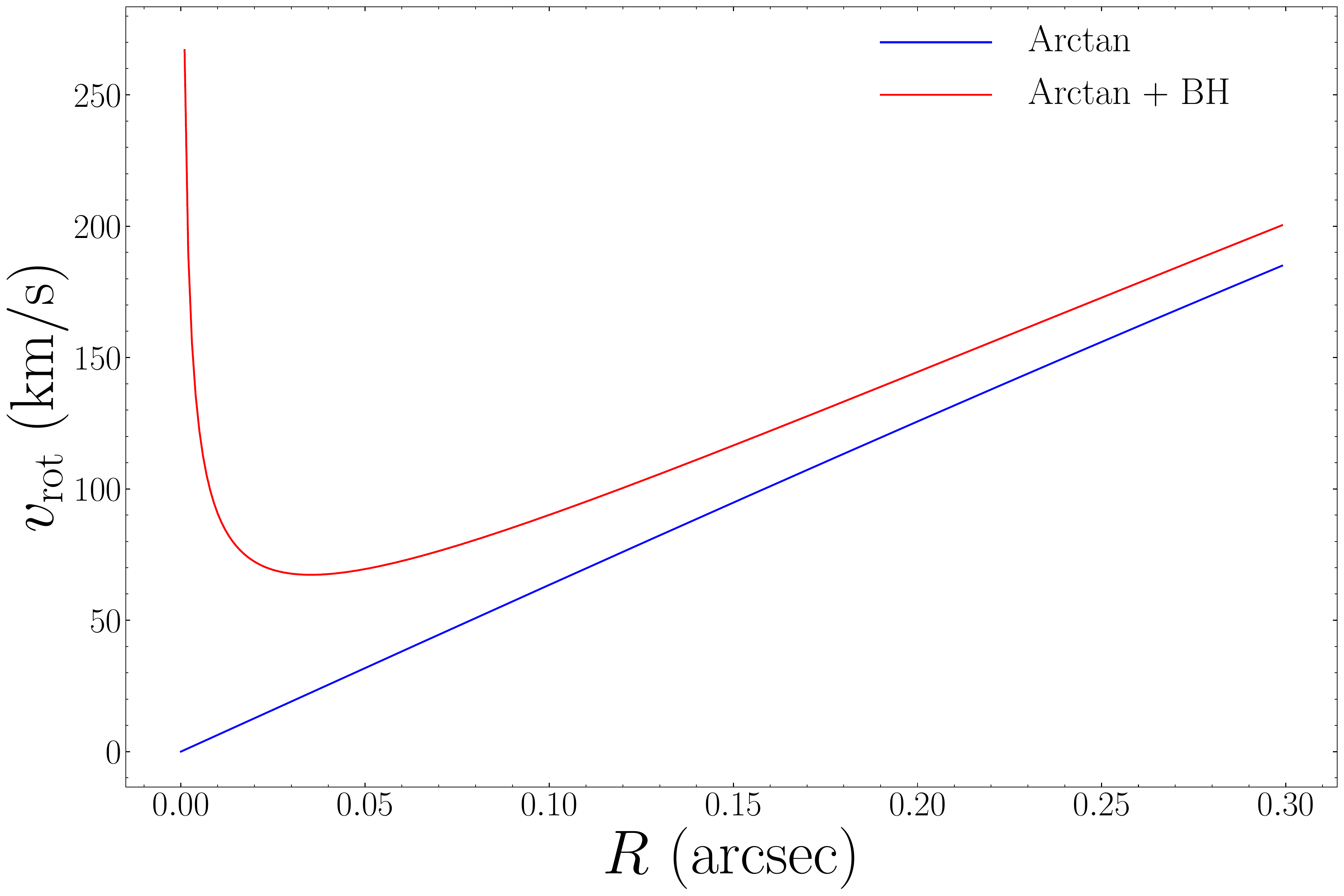}
    \caption{The two model rotation curves used in our KinMS analysis. The blue curve represents a `standard' arctan rotation curve, while the red adds a central point mass component (representing a black hole) with $M_{\rm BH}\,{=}\,2{\times}10^7\ M_{\odot}$.}
    \label{fig:rot-curve}
\end{figure}

We assume an exponential density profile for the gas, of the form $\Sigma\,{\propto}\,e^{-R/R_0}$, with a scale radius $R_0$\,=\,0\farcs15. For the velocity structure of these model disks, we consider two different rotation curves: (1) a standard rotation curve, $v_{\rm rot}$\,=\,(2\,$v_{\rm flat}$/$\pi$)\,arctan($R$), with $v_{\rm flat}\,{=}\,1,000$\ km\,s$^{-1}$, and (2) an additional central point-mass component, representing a supermassive black hole with mass $M_{\rm BH}\,{=}\,2{\times}10^7\ M_{\odot}$. $v_{\rm flat}$\ was chosen to reproduce the $\sim$150 km\,s$^{-1}$\ velocity difference between the two components in 0480's velocity profile, which we take as the maximum measured rotational velocity. The black hole mass was chosen assuming 0480 approximately follows the $M_{\rm BH}$--$\sigma_{\star}$\ relation \citep[e.g.,][]{gultekin2009}, with a $\sigma_{\star}$\,=\,124 km\,s$^{-1}$\ measured by SDSS. We show both of these rotation curves in Figure \ref{fig:rot-curve}. Lastly, we consider three different intrinsic gas velocity dispersions, $\sigma_{v}$, for the disks: (1) 25 km\,s$^{-1}$, much lower than observed, (2) 69 km\,s$^{-1}$, matching the average observed dispersion for 0480, and (3) an `intermediate' 50 km\,s$^{-1}$\ dispersion.

\begin{figure}[!h]
    \centering
    \includegraphics[width=0.87\linewidth]{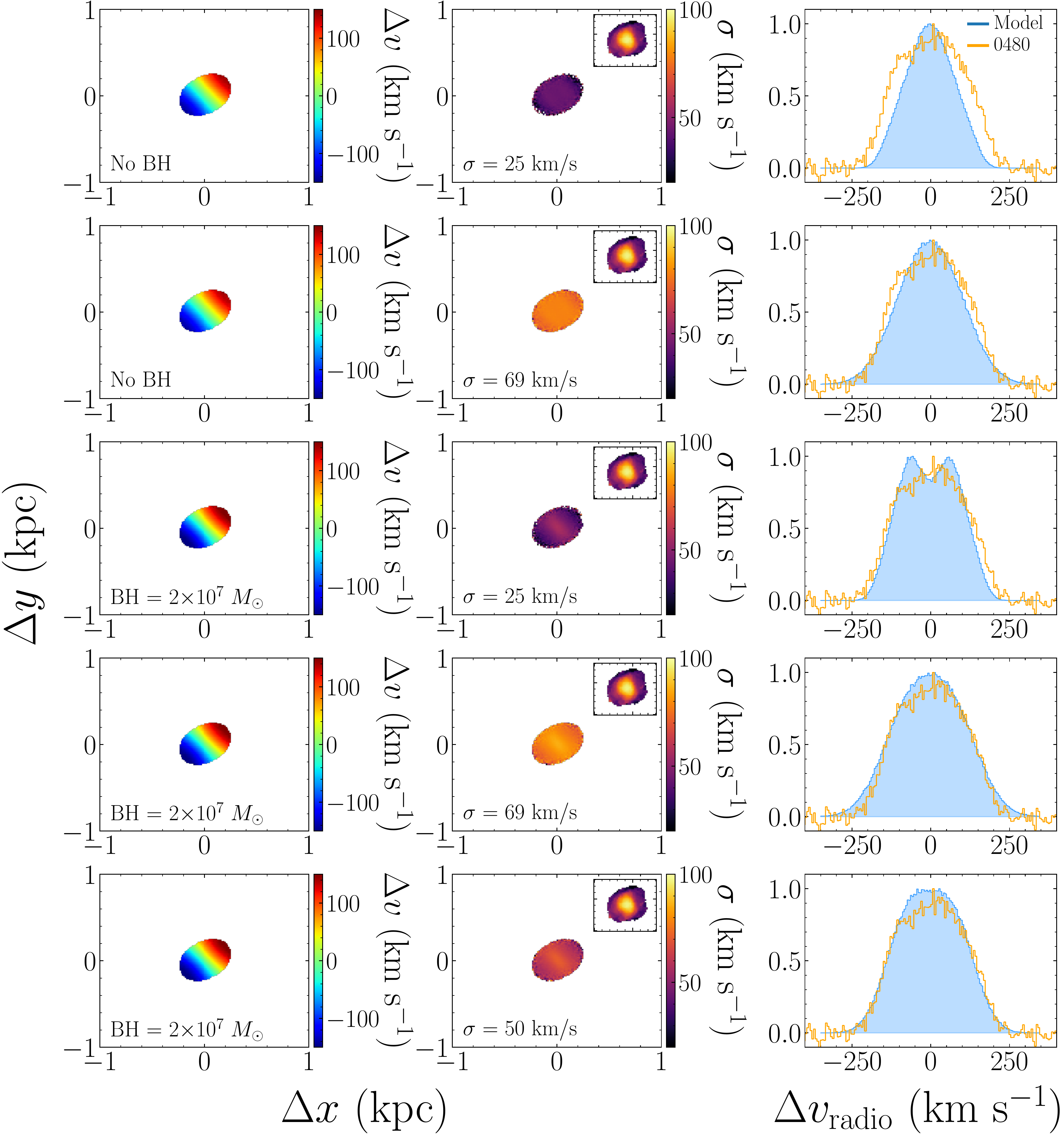}
    \caption{Moment 1 \&\ 2 maps and velocity profiles for each of the five model disks described in \textsc{Appendix} \ref{sec:beamsmear}. We also show the Moment 2 map of 0480 as an inset in the upper right of each panel in the second column. This is plotted on an identical color scale as the model Moment 2 maps (i.e.\ 20--100 km\,s$^{-1}$). Likewise, we show the velocity profile of 0480 (orange) overlaid on the extracted velocity profile for each model (blue).}
    \label{fig:beamsmear}
\end{figure}

We consider five distinct combinations of these rotation curves and intrinsic velocity dispersions: (1) a low-dispersion 25 km\,s$^{-1}$\ model, with a standard rotation curve and no central black hole, (2) a high-dispersion 69 km\,s$^{-1}$\ model, with a standard rotation curve and no central black hole, (3) a low-dispersion 25 km\,s$^{-1}$\ model, with a 2$\times$10$^7\ M_{\odot}$\ central black hole, (4) a high-dispersion 69 km\,s$^{-1}$\ model, with a 2$\times$10$^7\ M_{\odot}$\ central black hole, and (5) an intermediate-dispersion 50 km\,s$^{-1}$\ model, with a 2$\times$10$^7\ M_{\odot}$\ central black hole. In Figure \ref{fig:beamsmear} for each of these models we show the Moment 1 and 2 maps, as well as the extracted velocity profile. We show 0480's Moment 2 map and velocity profile for reference. It seems clear from the comparisons that a gas velocity dispersion higher than 25 km\,s$^{-1}$\ is required to reproduce 0480's Moment 2 map and the shape of its velocity profile, regardless of the presence of a central black hole. Of the five models, the intermediate velocity dispersion of 50 km\,s$^{-1}$\ with a black hole seems to best reproduce the data. 

Our data are not high enough resolution to fully model 0480's rotation curve and precisely determine the relative contributions of high inner rotation and intrinsic velocity dispersion. It is also possible that additional velocity components, such as compact bars or gas streamers, could also contribute to beam smearing of ordered motions. We can conclude from this analysis that, while beam smearing due to a steep central rotation curve may contribute significantly, it likely cannot account for more than $\sim$50\% of the velocity dispersion signal in the source with the clearest rotation. Given the high surface densities \textit{and} high velocity dispersions, beam smearing thus likely has little impact on our conclusions of high turbulent pressures (i.e.\ $\log_{10}P_{\rm turb}\,{>}\,8$) in the gas of these PSBs. 

\end{document}

%% file: obstbl.tex
\begin{deluxetable*}{ppppppppp}[t]
\setlength{\aboverulesep}{0pt}
\setlength{\belowrulesep}{-2pt}
\tablecaption{Observations\label{tab:obs}}
\tablecolumns{9}
\tabletypesize{\small}
\tablehead{%
\multicolumn{2}{c}{\hspace{20pt}Galaxy} &
\colhead{} &
\colhead{} &
\colhead{Cyc.\,3 (12\,m)} &
\colhead{Cyc.\,3 (7\,m)} &
\colhead{Cyc.\,4 (12\,m)} &
\colhead{Cont.\ RMS} &
\colhead{Line RMS} \\
\cmidrule(l{15pt}r{0pt}){1-2}
\colhead{(S18)} &
\colhead{(F15)} & 
\colhead{R.A.} &
\colhead{Decl.} &
\colhead{(min)} &
\colhead{(min)} &
\colhead{(min)} &
\colhead{(mJy/beam)} &
\colhead{(mJy/beam)} \\
\colhead{(1)} & 
\colhead{(2)} & 
\colhead{(3)} & 
\colhead{(4)} &
\colhead{(5)} & 
\colhead{(6)} & 
\colhead{(7)} &
\colhead{(8)} &
\colhead{(9)} 
}
\startdata
0379\_579\_51789 & S15 & 22:55:06.80 & +00:58:39.9 & 28 & 87 & \ldots & 0.033 & 0.94 \\
0480\_580\_51989 & H08 & 09:48:18.68 & +02:30:04.2 & 9.0 & \ldots & 31 & 0.026 & 0.62 \\
0570\_537\_52266 & S05 & 09:44:26.96 & +04:29:56.8 & 36 & 140 & 192 & 0.0091 & 0.44 \\
0637\_584\_52174 & S14 & 21:05:08.67 & $-$05:23:59.4 & 18 & 54 & \ldots & 0.044 & 1.4 \\
2360\_167\_53728 & H02 & 09:26:19.29 & +18:40:41.0 & 4.5 & \ldots & 9.0 & 0.036 & 1.2 \\
2777\_258\_54554 & H03 & 14:48:16.05 & +17:33:05.9 & 7.5 & 28 & \ldots & 0.071 & 2.0 \\
\enddata
\tablecomments{(1)--(2) Galaxy ID in SDSS Plate\_Fiber\_MJD notation, following \cite{smercina2018}, as well as the E+A ID assigned in \cite{french2015}. (3)--(4) Right Ascension and Declination. Total observation time, across all execution blocks with the: (5) 12\,m array in Cycle 3, (6) 7\,m array in Cycle 3, and (7) 12\,m array in Cycle 4. (8) RMS of the continuum image. (9) Line RMS in each 5.4 km\,s$^{-1}$\ channel.}
\vspace{-20pt}
\end{deluxetable*}

%% file: restbl.tex
\begin{deluxetable*}{ssssssssssss}[t]
\setlength{\aboverulesep}{0pt}
\setlength{\belowrulesep}{-2pt}
\tablecaption{Properties of the PSB Sample\label{tab:res}}
\tablecolumns{12}
\tablewidth{250pt}
\tablehead{%
\colhead{Galaxy} &
\colhead{} & 
\colhead{$D_{\rm L}$} & 
\colhead{$\log_{10}L_{\TIR}$} &
\colhead{$D_{\rm beam}$} &
\colhead{$L^{\prime}_{\rm CO(2-1)}$} &
\colhead{$L^{\prime,{\rm IRAM}}_{\rm 2-1}$} &
\colhead{$M_{\rm Mol}$} &
\colhead{$f_{\nu}{\rm (1.3\,mm)}$} &
\colhead{$f_{\nu,1.3}^{\rm obs}$} &
\colhead{$\theta_{\rm a} \times \theta_{\rm b}$} & 
\colhead{$f_{\rm core}$} \\ \cmidrule(l{7pt}r{7pt}){7-7} \cmidrule(l{3pt}r{3pt}){10-10}
\colhead{(S18)} &
\colhead{$z$} &
\colhead{(Mpc)} & 
\colhead{($L_{\odot}$)} &
\colhead{(kpc)} & 
\colhead{(10$^7$\ K km\,s$^{-1}$\ pc$^{2}$)} &
\colhead{$L^{\prime,{\rm ALMA}}_{\rm 2-1}$} & 
\colhead{(10$^8$\ $M_{\odot}$)} &
\colhead{(mJy)} &
\colhead{$f_{\nu,1.3}^{\rm SED}$} &
\colhead{(\arcsec)} &
\colhead{(\%)} \\
\colhead{(1)} & 
\colhead{(2)} & 
\colhead{(3)} & 
\colhead{(4)} & 
\colhead{(5)} &
\colhead{(6)} &
\colhead{(7)} & 
\colhead{(8)} &
\colhead{(9)} & 
\colhead{(10)} &
\colhead{(11)} & 
\colhead{(12)}
}
\startdata
0379 & 0.053 & 247.2 & 9.74 & 0.95 & 4.45\,$\pm$\,0.16 & 5.76\,$\pm$\,1.24 & 2.7 & $<$\,0.25 & $<$1.1 & $0.91 \times 0.85$ & 81 \\
0480 & 0.060 & 280.9 & 11.18 & 0.22 & 39.6\,$\pm$\,0.95 & $<$\,0.83 & 27.7 & 7.54\,$\pm$\,0.22 & 3.7 & $0.20 \times 0.17$ & 95 \\
0570 & 0.047 & 215.3 & 9.70 & 0.23 & 5.77\,$\pm$\,0.89 & $<$\,2.67 & 2.0 & $<$\,1.66 & $<$7.1 & $0.29 \times 0.20$ & 46 \\
0637 & 0.083 & 390.2 & 10.24 & 1.42 & 20.0\,$\pm$\,2.1 & 3.25\,$\pm$\,0.91 & 9.3 & $<$\,0.45 & $<$1.6 & $0.99 \times 0.78$ & 63 \\
2360 & 0.054 & 250.8 & 10.51 & 0.30 & 25.0\,$\pm$\,4.6 & 1.41\,$\pm$\,0.53 & 14.7 & $<$\,0.69 & $<$3.6 & $0.29 \times 0.26$ & 80 \\
2777 & 0.045 & 206.7 & 10.86 & 0.55 & 147\,$\pm$\,51 & 0.67\,$\pm$\,0.24 & 69.4 & 1.00\,$\pm$\,0.35 & 0.8 & $0.75 \times 0.48$ & 64 \\
\enddata
\tablecomments{(1) Galaxy ID, given as the SDSS plate number following \cite{smercina2018}. (2)--(3) The optical redshift, $z$, measured from the SDSS spectrum, and corresponding luminosity distance, $D_{\rm L}$, adopted throughout this paper, assuming a \cite{planck-cosmo} cosmology. (4) Total infrared (TIR) luminosity, from \cite{smercina2018} SED fitting. (5) Physical resolution of the elliptical beam, averaged over both axes. (6) Integrated CO(2--1) line luminosity from fits to the velocity profiles. Uncertainties are assessed as the standard deviation of 10$^4$\ bootstrap fits to each profile. (7) Ratio of integrated CO(2--1) line luminosity measured from IRAM \citep{french2015} and ALMA (this work) observations. Formal uncertainties were incorporated from both measurements; $<$\ denotes an upper limit. Some sources' IRAM measurements may be more uncertain than their formal errors; see \S\,\ref{subsec:mmol} for discussion. (8) Total molecular gas mass estimated from $L^{\prime}_{\rm CO(2-1)}$, assuming a Galactic CO-to-H$_2$\ conversion factor of $\alpha_{\rm CO}\,{=}\,4.35$\ (K km s$^{-1}$\ pc$^{2}$)$^{-1}$\ and a $^{12}$CO(2--1)/CO(1--0) line ratio of $R_{21}\,{=}\,0.59$. (9) Measured 1.3\,mm continuum flux density; $<$\ denotes an upper limit. (10) Ratio of observed 1.3\,mm flux density to the expectation from the best-fit total infrared \cite{draineli2007} model SED \citep{smercina2018}. (11) FWHM along major and minor axes of 2-D elliptical Gaussian fits to the CO(2--1) moment 0 maps. All sources' FWHMs correspond to the respective beam widths, i.e.\ all six galaxies contain unresolved CO `cores'. (12) Fraction of CO(2--1) emission within 3$\sigma$\ of the center --- i.e. fraction in the unresolved `core' (expressed as a percent). Here, $\sigma$\ is used relative to the 2-dimensional Gaussian fit, as the radius containing 68\% of the light.}
\vspace{-20pt}
\end{deluxetable*}

%% file: denstbl.tex
\begin{deluxetable}{bbbbbbbb}[!h]
\setlength{\aboverulesep}{0pt}
\setlength{\belowrulesep}{-2pt}
\tablecaption{Figure Data\label{tab:dens}}
\tablecolumns{8}
\tablewidth{250pt}
\tablehead{%
\colhead{Galaxy} &
\colhead{$R_{50,\star}$} &
\colhead{$R_{\rm 50,CO}$} &
\colhead{$\sigma_v$} &
\colhead{$\log_{10}\,\Sigma_{\rm Mol}$} & 
\colhead{$\log_{10}\,P_{\rm turb}$} &
\colhead{$\log_{10}\,\Sigma_{\rm SFR,low}$} & 
\colhead{$\log_{10}\,\Sigma_{\rm SFR,high}$} \\
\colhead{(S18)} &
\colhead{(kpc)} &
\colhead{(kpc)} &
\colhead{(km\,s$^{-1}$)} &
\colhead{(M$_{\odot}$\,pc$^{-2}$)} & 
\colhead{(K\,cm$^{-3}$)} &
\colhead{(M$_{\odot}$\,yr$^{-1}$\,kpc$^{-2}$)} & 
\colhead{(M$_{\odot}$\,yr$^{-1}$\,kpc$^{-2}$)} \\
\colhead{(1)} &
\colhead{(2)} &
\colhead{(3)} &
\colhead{(4)} &
\colhead{(5)} & 
\colhead{(6)} &
\colhead{(7)} & 
\colhead{(8)}
}
\startdata
0379 & 2.95 & 0.95 & \ldots & 2.57 & \ldots & $-$0.59 & 0.10 \\
0480 & 1.66 & 0.22 & 69 & 4.85 & 9.8 & 1.59 & 2.78 \\
0570 & 2.61 & 0.23 & 112 & 3.67 & 9.1 & 0.37 & 1.05 \\
0637 & 4.57 & 1.43 & \ldots & 2.77 & \ldots & $-$0.37 & 0.21 \\
2360 & 1.55 & 0.30 & 47 & 4.32 & 8.8 & 1.18 & 1.86 \\
2777 & 3.08 & 0.56 & \ldots & 4.46 & \ldots & 1.15 & 1.71 \\
\enddata
\tablecomments{(1) Galaxy ID. (2) Stellar half-light radius, from SDSS photometry \citep{french2015,smercina2018}. (3) CO half-mass radius, estimated from comparing the Gaussian fits to the full gas distributions (see Figure \ref{fig:sizes} caption). (4) Average central velocity dispersion, used to calculate $P_{\rm turb}$. (5) Molecular gas surface density within the central core of each galaxy. (6) Turbulent pressure calculated from the measured velocity dispersion and molecular gas surface density, using Equation \ref{eq:3}. (7) Lowest estimate on SFR surface density from the three available infrared tracers. (8) Highest estimate on SFR surface density from the three available infrared tracers.}
\vspace{-50pt}
\end{deluxetable}